\documentclass[a4paper,fleqn,usenatbib,useAMS]{mnras}

\usepackage{color}
\usepackage{graphicx}
\usepackage{amsmath}	% Advanced maths commands
\usepackage{amssymb}	% Extra maths symbols
\usepackage{multicol}        % Multi-column entries in tables
\usepackage{bm}		% Bold maths symbols, including upright Greek
\usepackage{pdflscape}	% Landscape pages
\usepackage{subcaption}
\usepackage{comment}
\usepackage[T1]{fontenc}
\usepackage{ae,aecompl}
\usepackage{enumitem}
\usepackage{colortbl}

\usepackage{newtxtext,newtxmath}
\usepackage{etoolbox}
\makeatletter
\patchcmd\@combinedblfloats{\box\@outputbox}{\unvbox\@outputbox}{}{\errmessage{\noexpand patch failed}}
\makeatother

\title[Embedding globular clusters in dark matter minihalos]{Embedding globular clusters in dark matter minihalos solves the cusp-core and timing problems in the Fornax dwarf galaxy}

\author[P. Boldrini, R. Mohayaee, J. Silk]{Pierre Boldrini$^{1}$\thanks{Contact e-mail: \href{mailto:boldrini@iap.fr}{boldrini@iap.fr}}, {Roya Mohayaee$^{1}$}, {Joseph Silk$^{1,2,3}$}
\\
$^{1}$Sorbonne Universit\'e, CNRS, UMR 7095, Institut d'Astrophysique de Paris, 98 bis bd Arago, 75014 Paris, France\\
$^{2}$Department of Physics and Astronomy, The Johns Hopkins University, Baltimore MD 21218, USA\\
$^{3}$Beecroft Institute for Particle Astrophysics and Cosmology, Department of Physics, University of Oxford, Oxford OX1 3RH, UK}
\date{In original form 2019 July 16.}

\pubyear{2019}

\DeclareMathOperator{\erf}{erf}
\begin{document}
\label{firstpage}
\pagerange{\pageref{firstpage}--\pageref{lastpage}}
\maketitle

\begin{abstract}
We use a fully GPU $N$-body code to demonstrate that dark matter minihalos, as a new component of globular clusters, resolve both the timing and cusp-core problems in Fornax if the five (or six) globular clusters were recently accreted ($\leq$ 3 Gyr ago) by Fornax. Under these assumptions, infall of these globular clusters does not occur and no star clusters form in the centre of Fornax in accordance with observations. We find that crossings of globular clusters that have  DM minihalos near the Fornax centre induce a cusp-to-core transition of the dark matter halo and hence resolve the cusp-core problem in this dwarf galaxy. The dark matter core size depends on the frequency of globular cluster crossings. Our simulations clearly demonstrate also that between the passages, the dark matter halo can regenerate its cusp. Moreover, our models are in good agreement with constraints on the dark matter masses of globular clusters as our clusters  lose a large fraction of their initial dark matter minihalos. These results provide circumstantial evidence for the universal existence of dark matter halos in globular clusters.
\end{abstract}


%%%%%%%%%%%%%%%%%%%%%%%%%%%%%%%%%%%%%%%%%%%%%%%%%%

%%%%%%%%%%%%%%%%% BODY OF PAPER %%%%%%%%%%%%%%%%%%

% The MNRAS class isn't designed to include a table of contents, but for this document one is useful.
% I therefore have to do some kludging to make it work without masses of blank space.

\section{Introduction}

Globular clusters (GCs) are gravitationally bound groupings of mainly old stars, formed in the early phases of galaxy formation. The origin of GCs is one of the key unsolved astrophysical problems. There are also various unresolved open questions regarding the formation and evolution of GCs during galaxy formation and assembly within a cosmological framework \citep{2018RSPSA.47470616F}. Despite their relevance to star formation, galaxy evolution and cosmology, there is no clear consensus on the formation of GCs. One can classify all proposed scenarios for old GC formation into two broad categories. Firstly, GCs originated as gravitationally bound gas clouds in the early Universe and are formed inside their present-day host galaxies. This corresponds to a primary  in-situ formation process\citep{1968ApJ...154..891P,2005ApJ...623..650K,2015MNRAS.454.1658K}. Secondly, GCs can be formed around the time of reionization in dark matter (DM) minihalos that later merge to become a part of the present-day host galaxy. It corresponds to a secondary  galactic origin in a similar way to formation of  dwarf galaxies \citep{1984ApJ...277..470P,2002ApJ...566L...1B,2005ApJ...619..243M,2016ApJ...831..204R}. Little star formation can take place during the reionization epoch inside these DM minihalos with virial masses less than $10^8$ M$_{\sun}$, as the heated gas can escape \citep{1999ApJ...523...54B}. However, the reionization of the Universe can actually trigger star formation in DM minihalos through different reionization-regulated positive feedback mechanisms \citep{2001ApJ...560..592C,2002ApJ...575...49R}. While most GCs are old, there are also young GCs such as Terzan 7 and Whiting 1 \citep{2010ApJ...718.1128L,2017A&A...598L...9M}, as well as the populations of young globular clusters in the MagellanIc Clouds and many other nearby star-forming galaxiers. These GCs could be formed via a different mechanism from that of  the old GCs. Here, we focus on the second scenario where GCs are formed at the centres of their own DM minihalos. 

Positions and proper motions for GCs in the Galaxy from the second data release of the Gaia mission will offer unique insights into GC dynamics \citep{2018A&A...616A..12G}. Until now, these DM minihalos have not been detected, but there are GC features and perhaps observational signatures \citep{2016MNRAS.462.1937S}. For instance, an extended diffuse stellar envelope can highlight whether GCs might be embedded in DM minihalos \citep{2009AJ....138.1570O,2016MNRAS.461.3639K,2017MNRAS.471L..31P}. Also, GCs orbiting in the inner regions of the Galaxy may lose a large fraction of any DM minihalo mass \citep{2002ApJ...566L...1B}.

Dwarf spheroidal (dSph) galaxies are among the most dark matter-dominated galaxies in the Universe \citep{2013NewAR..57...52B,2013pss5.book.1039W}, and their dynamics are determined entirely by their DM halos. Fornax, the most massive of the Milky Way dSphs, is the only one to have five (or perhaps six) GCs in orbit \citep{2019ApJ...875L..13W}. Dynamical friction should have entrained the infall of these clusters within a few Gyr due  to the dense background of dark matter in this galaxy \citep{1943ApJ....97..255C,1975ApJ...196..407T,1976ApJ...203..345T}. However, there is no nuclear star cluster (NSC) in Fornax, where we are still observing orbiting GCs. This has become known as the Fornax timing problem \citep{2000ApJ...531..727O}. Many $N$-body simulations have been performed to study the Fornax timing problem \citep{2000ApJ...531..727O,2006MNRAS.368.1073G,2006MNRAS.373.1451R,2009MNRAS.399.1275P,2012MNRAS.426..601C}. Recent simulations of the Fornax dwarf galaxy found that the observed GCs are not necessarily incompatible with a Navarro-Frenk-White (NFW) DM profile and that this depends strongly on GC initial conditions \citep{2019MNRAS.485.2546B}. However Fornax GCs remain a puzzle for understanding their survival over a Hubble time.

Even if the CDM theory can successfully explain various observations at different scales, there are unresolved problems concerning the DM halo profile in galaxies such as the cusp-core and diversity problems. The first one depicts the discrepancy between observations, which generally reveal cores at the centres of dwarf galaxies \citep{1994Natur.370..629M,1995ApJ...447L..25B,2001ApJ...552L..23D,2003ApJ...583..732S,2005AJ....129.2119S,2011ApJ...742...20W}, and cosmological simulations, which generally predict a cuspy profile \citep{1997ApJ...490..493N,1997ApJ...477L...9F,1998ApJ...499L...5M,2010MNRAS.402...21N}. Stellar kinematics and the survival of Fornax GC systems  suggest that Fornax has a dark matter core \citep{2006MNRAS.368.1073G,2011MNRAS.411.2118A,2011ApJ...742...20W,2012MNRAS.426..601C,2019MNRAS.484.1401R}. In order to resolve this discrepancy, many mechanisms involving baryons have been proposed, which could transform cusps into cores via time variations in the gravitational potential caused by stellar feedback redistributing massive gas clouds, thereby generating bulk motions and galactic winds along with DM heating by dynamical friction of massive clumps \citep[e.g.][]{1996ApJ...462..563N,2011ApJ...736L...2O,2013MNRAS.429.3068T,2012MNRAS.421.3464P,2001ApJ...560..636E,2010ApJ...725.1707G,2011MNRAS.418.2527I}.

In this work, we embed GCs in DM minihalos in order to simultaneously resolve the timing and the cusp-core problems in Fornax with the most prevalent initial conditions from the Illustris cosmological simulation \citep{2018MNRAS.473.4077P}. Our $N$-body simulations performed on GPU allow us to study the evolution of the DM density profile in the centre of Fornax at high resolution. The paper is organized as follows. Section 2 provides a description of the Fornax system and the $N$-body modelling. In Section 3, we outline details of our numerical simulations. In Section 4, we illustrate our simulation results and discuss the implications of DM minihalos of GCs on the timing and cusp-core problems. Section 5 presents our conclusions.

\section{Fornax-globular cluster system}

\begin{figure}
\centering
\includegraphics[width=0.47\textwidth]{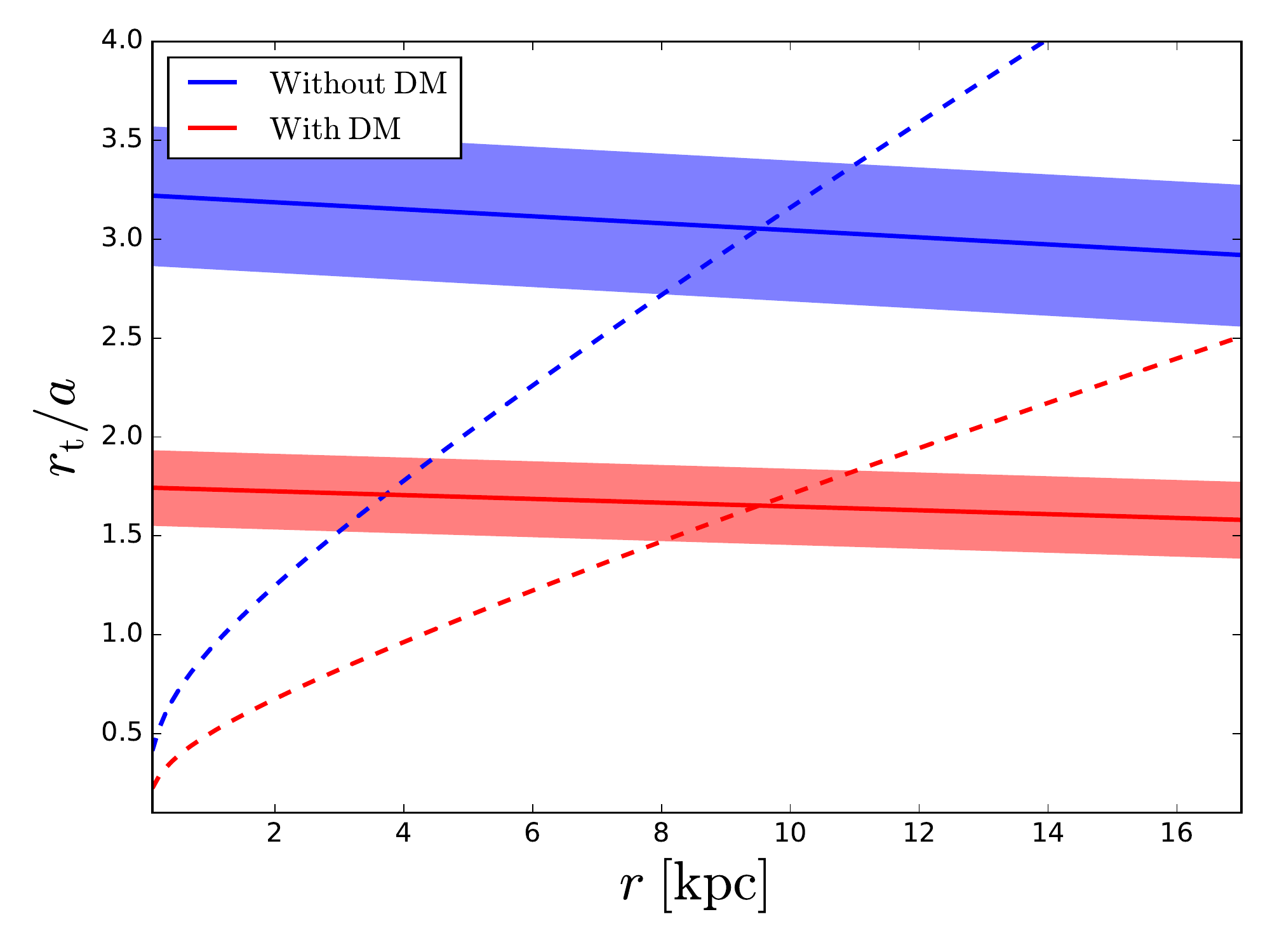}
\caption{{\it Tidal impact of MWG:} Ratio between the tidal radius and the typical size of GCs as a function of the orbital radius centered on Fornax for GCs with and without a DM minihalo. For GCs composed only of stars, we assume that $a=r_{t_{max}}$, where $r_{t_{max}}$ is the highest initial tidal radius of our GCs in Fornax. However, we set $a=r_{vir}$, where $r_{vir}$ is the virial radius of the $2\times10^7$ M$_{\sun}$ minihalo for GCs embedded in a DM minihalo. Solid (dashed) line represents the influence of the MWG (Fornax) on GCs. For the MW, the error bands are due to the uncertainty on pericentre and apocentre of Fornax \citep{2018A&A...616A..12G}. The MWG has no major impact on Fornax GCs, because the tidal radius is always higher than the typical size of the GCs at any radii. This is the reason why we assume that tidal disruptions are mostly driven by Fornax and that our dSph model is in isolation.}
\label{fgr99}
\end{figure}

The dSph galaxy Fornax is one of the more dark matter-rich satellites of the Milky Way Galaxy  with a stellar mass of about $10^{8}$ M$_{\sun}$ at a distance of around 147 kpc \citep{2016A&A...590A..35D}. We consider only the first five GCs with masses of about $10^{5}$ M$_{\sun}$ and average projected distances of about 1 kpc. In this section, we present the models for Fornax and its GCs that provide the initial conditions for our simulations. A live gravitational system is necessary to capture tidal stripping and dynamical friction. Thus, Fornax has to be modelled as a live galaxy with its GCs, i.e. a self-gravitating system composed of star and DM particles.  

It has been shown that simulated GCs formed within Fornax, modelled with a cuspy or cored DM halo, are compatible with observations \citep{2019MNRAS.485.2546B}. However, the origin of the Fornax DM core remains unsolved. Here, we consider two scenarios to explain the timing problem. We assume that GCs have experienced either early accretion 10-12 Gyr ago (z=3) or  recent accretion 2-4 Gyr ago (z=0.36) by the Fornax galaxy. Fornax GCs are all dominated by an old population (>10 Gyr), which gives  an uncertainty in the GC age determinations \citep{2016A&A...590A..35D}.

\subsection{Fornax galaxy}

We construct Fornax as a live galaxy composed of stars and DM particles only, since dSph galaxies contain little or no gas today. The Fornax stellar component is modelled, due to the presence of  a stellar core ($\mathrm{r_{0}=0.668\;kpc}$ \citep{2010MNRAS.408.2364S}), by a Plummer profile \citep{1911MNRAS..71..460P}:
\begin{equation}
\rho(r)=\frac{3a^{2}M_{0}}{4\pi}(r^{2}+a^{2})^{-5/2},
\label{plu}
\end{equation}
where $a=r_{0}(\sqrt{2}-1)^{-1/2}$ and $M_{0}=3.82\times10^{7}$ $M_{\sun}$ are the scale parameter and the mass, respectively \citep{2016A&A...590A..35D}. For the DM halo of Fornax, we assume a NFW form \citep{1996ApJ...462..563N}: 
\begin{equation}
\rho(r) = \rho_{0}\left(\frac{r}{r_{\mathrm{s}}}\right)^{-1}\left(1+\frac{r}{r_{\mathrm{s}}}\right)^{-2},
\label{eq1}
\end{equation}
with the central density $\rho_{0}$ and scale radius $r_{\mathrm{s}}$. For the simulations, we consider a Fornax-like dSph with mass of $10^9$ M$_{\sun}$ at redshift $z$ depending on the considered scenario. Given the halo mass and redshift, the scale radius $r_{\mathrm{s}}$ of the Fornax halo was estimated from cosmological $N$-body simulations \citep{2012MNRAS.423.3018P}.

\subsection{Globular clusters}

\begin{table}
\begin{center}
\label{tab:landscape}
\begin{tabular}{cccccccccc}
 \hline
   Redshift &Object & $r$ &  $v_x$ & $v_y$ & $v_z$ & $|v|$ \\
    & & [kpc] & [km/s] & [km/s] & [km/s] & [km/s] \\
    \hline
   $z=3$ &$E_1$ & 2.11 & 20.22 & 1.6 & 8.1 & 21.8 \\
    &$E_2$ & 2.74 & 16.42 & 3.28 & 15.75 & 22.99 \\
    &$E_3$ & 1.1 & 38.62 & 15.44 & 18.94 & 39.97 \\
    &$E_4$ & 1.76 & 13.46 & 21.46 & 26.87 & 36.94\\
    &$E_5$ & 1.14 & 32.81 & 6.93 & 7.89 & 34.37\\
    \hline
   $z=0.36$ &$O_1$ & 5.32 & 13.9 & 1.31 & 14.38 & 20.04 \\
    &$O_2$ & 2.07 & 9.43 & 19.05 & 21.42 & 30.18 \\
    &$O_3$ & 1.95 & 37.49 & 4.55 & 6.21 & 38.28 \\
    &$O_4$ & 2.19 & 3.51 & 3.87 & 34.0 & 34.39\\
    &$O_5$ & 2.05 & 19.62 & 29.8 & 15.1 & 38.75\\
    \hline
\end{tabular}
\caption{Initial conditions for five GCs $E_{i}$ at z=3 ($\simeq$ 12 Gyr) and five GCs $O_{i}$ at z=0.36 ($\simeq$ 4 Gyr), determined by using Illustris TNG-100 data. We consider two scenarios to explain the timing problem. We assume that GCs have experienced early accretion onto the Fornax galaxy 10-12 Gyr ago, or alternatively have undergone recent accretion 4 Gyr ago (z=0.36). This is the reason why we determine the most prevalent positions and velocities of 7.5$\times10^{6}$ M$_{\sun}$ particles at these two redshifts, which have the same projected distances $D_{\mathrm{obs}}$ as Fornax GCs at z=0 using multidimensional binned statistics. The Illustris DM mass resolution is of the same order as our minihalo mass.}
\label{tab1}
\end{center}
\end{table}

In our model, GCs with a stellar mass $M_{*}$ are formed at the centre of DM minihalos with a virial mass $M_{\rm mh}$. We fix the mass ratio $M_{*}/M_{\rm mh}$ to 0.05. For the five surviving GCs orbiting in Fornax, we assume a two-component model that includes a stellar component and a DM minihalo:
\begin{equation}
M_{\rm model}(r) = M_{*}(r) + M_{\rm mh}(r).
\end{equation}
Our $N$-body realizations assume a \cite{1962AJ.....67..471K} stellar density distribution,
\begin{equation}
\rho(r) = \rho_{0}\left[\left(1+\left(\frac{r}{r_{\rm k}}\right)^2\right)^{-1/2} - C\right], \quad C=\left[1+\left(\frac{r_{\rm t}}{r_{\rm k}}\right)^2\right]^{-1/2},
\end{equation}
where $r_{\rm k}$ and $r_{\rm t}$ are the King and tidal radii, respectively. For all the simulations, we chose a King radius $r_{\rm k} = 1$ pc lower than the observed radius, because it is susceptible to increase through dynamical processes such as mass loss \citep{2003MNRAS.340..175M}. 
The DM minihalos are distributed in a spherical halo following a NFW profile with a mass of $2\times10^7$ M$_{\sun}$ (see Equation~\ref{eq1}) at redshift $z=3$ or $z=0.36$ depending on the accretion scenario. Given the halo mass and redshift, the halo concentration $c_{200}$ of the DM minihalos can be estimated from cosmological $N$-body simulations \citep{2012MNRAS.423.3018P}.

We wish to provide relevant GC initial conditions from cosmological simulations. We found 7395 isolated subhalos with mass of about $10^9$ M$_{\sun}$ in the Illustris TNG-100 simulation \citep{2018MNRAS.473.4077P}. The mass resolution of TNG-100 is 7.5$\times10^{6}$ M$_{\sun}$, which is of the same order as our minihalo mass. The current projected distances $D_{\mathrm{obs}}$ for the GCs are from 0.24 to 1.6 kpc, which is the minimum distance between GCs and the Fornax centre. In order to determine the GC positions and velocities at these redshifts, we select particles of the isolated subhalos at z=0, which have the same projected distances as Fornax GCs at z=0 and we find corresponding particles at redshift z=3 and z=0.36. We subsequently compute multidimensional binned statistics for this 6-dimensional space in order to determine the maximum weights for each dimension. We applied  Scott's rule to determine the optimal bin width given by 
\begin{equation}
    \Delta_{\rm b}=\frac{3.5\sigma}{n^{1/3}},
\end{equation}
where $\sigma$ is the standard deviation of one dimension, and $n$ is the number of points. Our initial conditions for GCs, positions and velocities, calculated with this method are listed in Table~\ref{tab1}. 

\subsection{Milky Way Galaxy tidal field}

As Fornax is a galaxy satellite of the Milky Way Galaxy  (MWG), we may expect that the MW tidal field to have an impact on GCs. We estimated this by calculating the theoretical tidal radius of our GCs with and without a DM minihalo in the simulation \citep{2010MNRAS.401..105K}. The tidal radius marks the distance beyond which stars can escape the GC. For our calculation, we assume that the MWG potential, based on the model of \cite{1991RMxAA..22..255A}, consists of a stellar bulge as a Plummer sphere \citep{1911MNRAS..71..460P}, a disc represented by the potential from \cite{1975PASJ...27..533M} and a spherical DM halo described by a NFW profile \citep{1997ApJ...490..493N}. For this model, we used the revised parameters from \cite{2013A&A...549A.137I} (see their Table 1). The tidal radius is calculated at the mean radius, because it is currently overestimated at the pericentre. Indeed, the variation of the tidal field over time is sufficiently fast that the cluster cannot adapt to the changing environment, but rather behaves as if it experiences a single mean tidal field along its orbit. Fig.~\ref{fgr99} describes the ratio between the tidal radius and typical size of GCs as a function of the orbital radius centered on Fornax of GCs with and without a DM minihalo. We assume that the size of a GC (with a DM minihalo) corresponds to its initial tidal (virial) radius in Fornax. For GCs composed only of stars, we consider the highest initial tidal radius $r_{t_{max}}$ of our GCs in Fornax as the typical size $a$. However, we set the virial radius $r_{vir}$ of the $2\times10^7$ M$_{\sun}$ minihalo as the typical size for GCs embedded in a DM minihalo. For the MWG, the error bands are due to the uncertainty in pericentre and apocentre of Fornax \citep{2018A&A...616A..12G}. We showed that the MWG (solid line) has no major impact on Fornax GCs, because the tidal radius is always higher than the typical size of the GCs at any radii. On the contrary, Fornax (dashed line) will give rise to tidal disruptions of the GCs, especially those with DM minihalos. Finally, we assume that this dynamical process is mostly driven by Fornax and that our dSph model is in isolation.

\section{$N$-body simulations}

To generate our live objects, we use the initial condition code \textsc{magi}. Adopting a distribution-function-based method, it ensures that the final realization of the galaxy is in dynamical equilibrium \citep{2017arXiv171208760M}. We perform our simulations with the high performance collisionless $N$-body code \textsc{gothic}. This gravitational octree code runs entirely on GPU and is accelerated by the use of hierarchical time steps in which a group of particles has the same time step \citep{2017NewA...52...65M}. We run $N$-body simulations with five GCs orbiting in Fornax galaxy by adopting a softening length of $\mathrm{\epsilon_0=1\;pc}$ and an accuracy control parameter of $A_0=2^{-7}$. Considering a softening length similar to the King radius maintain the dynamical stability of an isolated GC (see Fig.s~\ref{app21}. We assess the impact of the softening length $\epsilon$ on the stellar density profile for an isolated GC with and without a DM minihalo, and  the orbital decay and the mass loss of GCs and the evolution of the DM profile of Fornax (see in Fig.s~\ref{app21},~\ref{app1} ~\ref{app12} and ~\ref{app2}). We consider both  early and recent accretion of GCs by Fornax. For these two scenarios, we run simulations with five GCs embedded in their own DM minihalo and five GCs composed only of stars as usual. Initially, GCs with and without DM minihalo have the same most prevalent positions and velocities (see in Table~\ref{tab1}). The halo and the stellar components of Fornax are represented by $N=10^{7}$ particles. The mass ratios for these two components determine the particle number for each component. DM minihalos and GC stars are represented by about $10^{5}$ and $10^{4}$ particles, respectively. In fact, we require that the particle masses of all components are set to be equal in order to reduce numerical artifacts. To analyze our data and extract our results, we use a \textsc{python} module toolbox, \textsc{pnbody 4.0} \citep{2013ascl.soft02004R}.

\section{Results}

Next, we present and discuss our simulation results. To describe cluster orbital decay, we calculate the distance between the cluster and Fornax mass centres at each snapshot, in order to get the orbital radius. To estimate the GC mass loss, we count only bound particles. The main challenge with the Fornax timing problem is to find a way to delay  GC orbital decay in order to observe GCs in orbit today instead of a nuclear star cluster at the centre of the galaxy. The GC infall is due to dynamical friction generated by the Fornax DM field. Compared to stellar GCs, DM minihalos are expected to fall more rapidly towards the galaxy centre due to their higher masses. We consider two scenarios for Fornax GCs, as described in the following sections.  

\subsection{Early accretion}

\begin{figure}
\centering
\includegraphics[width=0.47\textwidth]{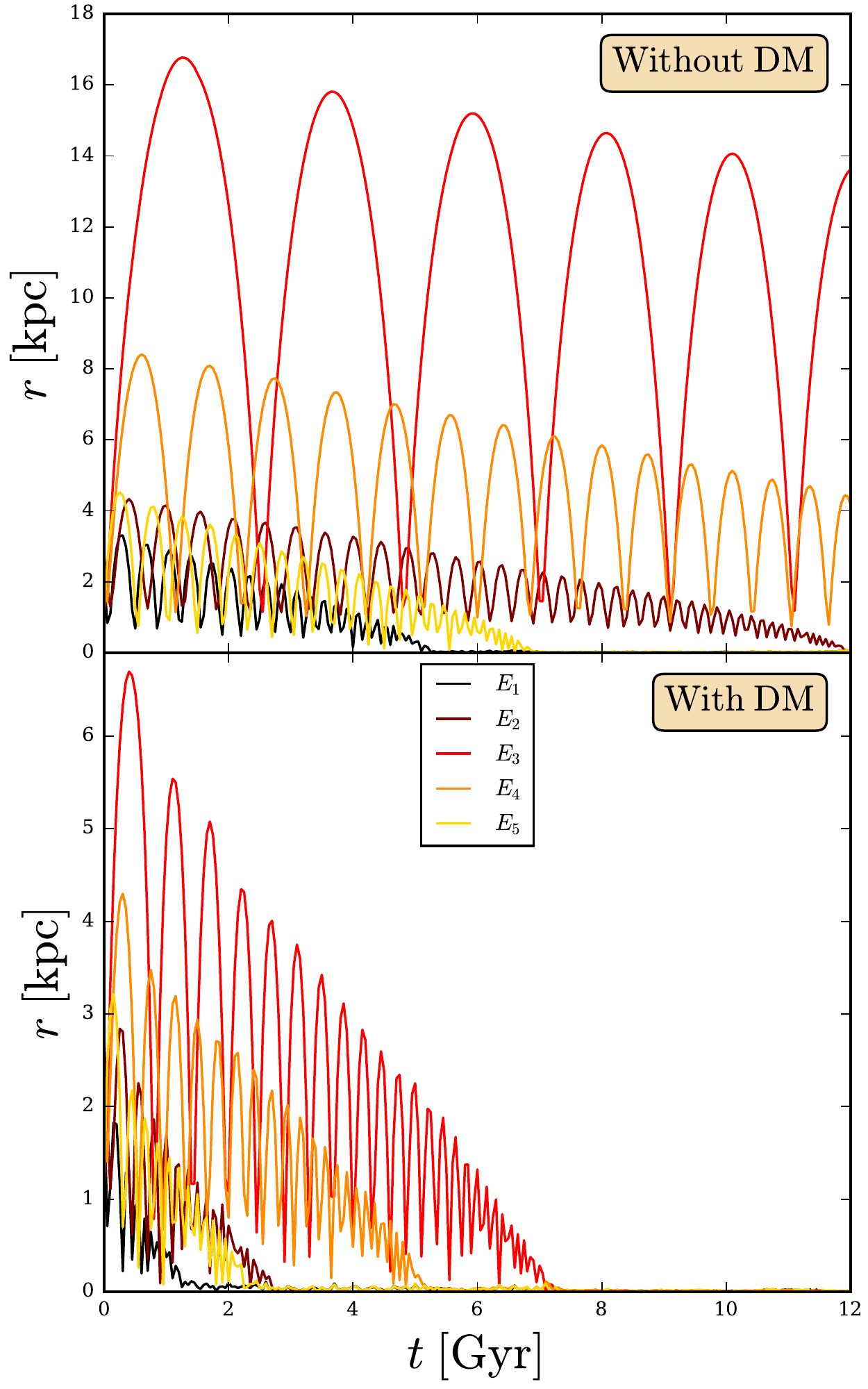}
\caption{{\it DM minihalos accelerate  GC infall:} Orbital decay of the 5 GCs without ({\it upper})
and with ({\it lower panel}) in a $2\times10^7$ M$_{\sun}$ DM minihalo over 12 Gyr because the Fornax GCs are all dominated by ancient (>10 Gyr) populations of stars \citep{2016A&A...590A..35D}. These radii correspond to the distances between each GC mass centre and Fornax mass centre. Initially, $10^6$ M$_{\sun}$ GCs with and without DM minihalos have the same most prevalent positions and velocities at $z=3$ from Illustris TNG-100 cosmological simulations (see objects $E_i$ in Table~\ref{tab1}). Our initial conditions entail an accretion process by the Fornax galaxy with eccentric orbits for all the GCs. The infall of GCs with and without a DM minihalo rules out the early accretion scenario for Fornax GCs as this dSph has five orbiting GCs observed at the present day.}
\label{fgr2a}
\end{figure}

\begin{figure}
\centering
\includegraphics[width=0.47\textwidth]{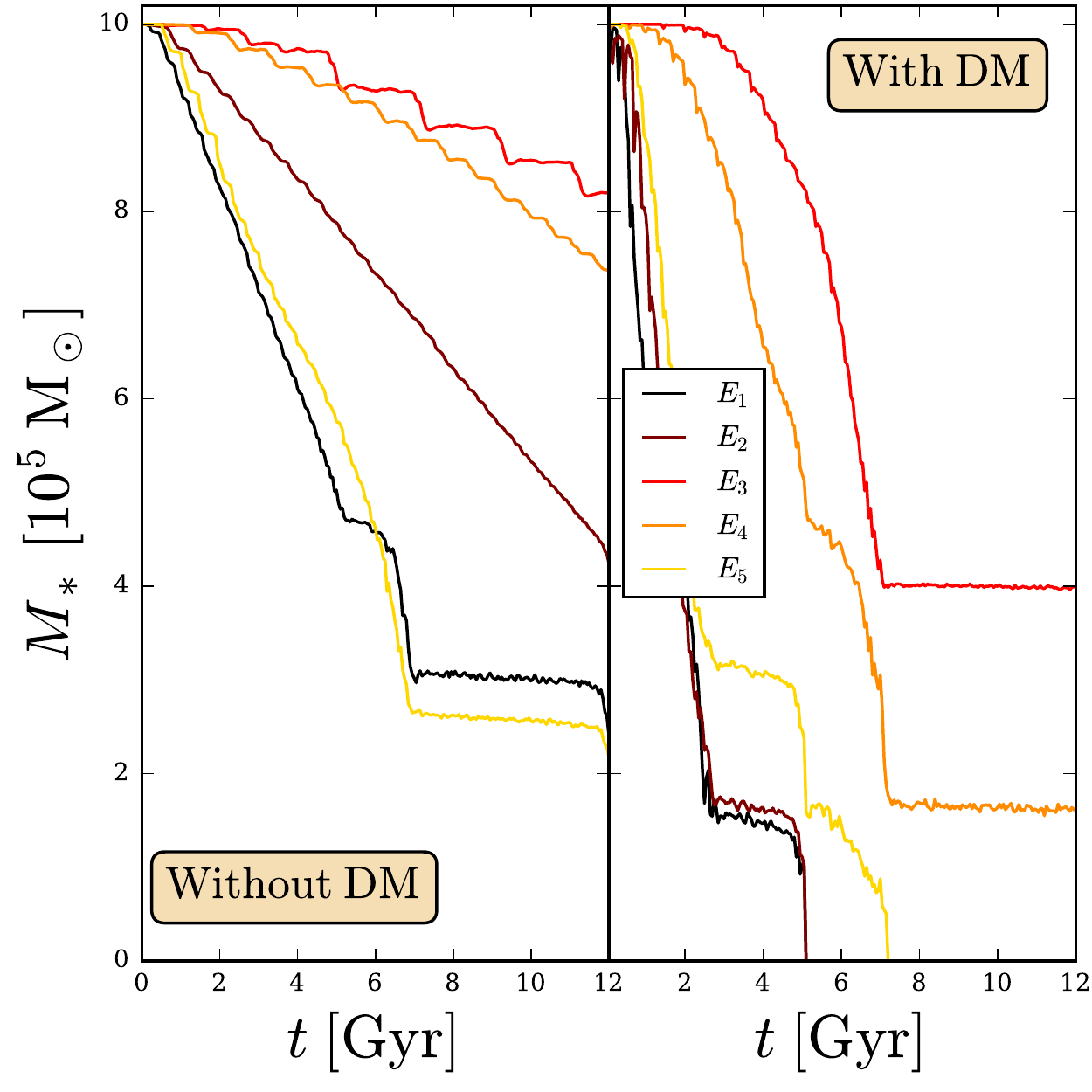}
\caption{{\it DM minihalos accelerate tidal stripping:} Evolution of the mass loss of the GC stellar component for the 5 GCs without ({\it left panel}) and with ({\it right panel}) a $2\times10^7$ M$_{\sun}$ DM minihalo over 12 Gyr. The initial  stellar masses of GCs are $10^6$ M$_{\sun}$. In order to estimate the GC mass loss, we count only bound particles. GCs with a $2\times10^7$ M$_{\sun}$ DM minihalo lost more stars compared to stellar GCs, because minihalos induce  major tidal stripping. All stellar GCs survive in this early accretion scenario. However, we observe that three GCs with a DM minihalo are completely tidally stripped within 8 Gyr.}
\label{fgrb}
\end{figure}

\begin{figure}
\centering
\includegraphics[width=0.47\textwidth]{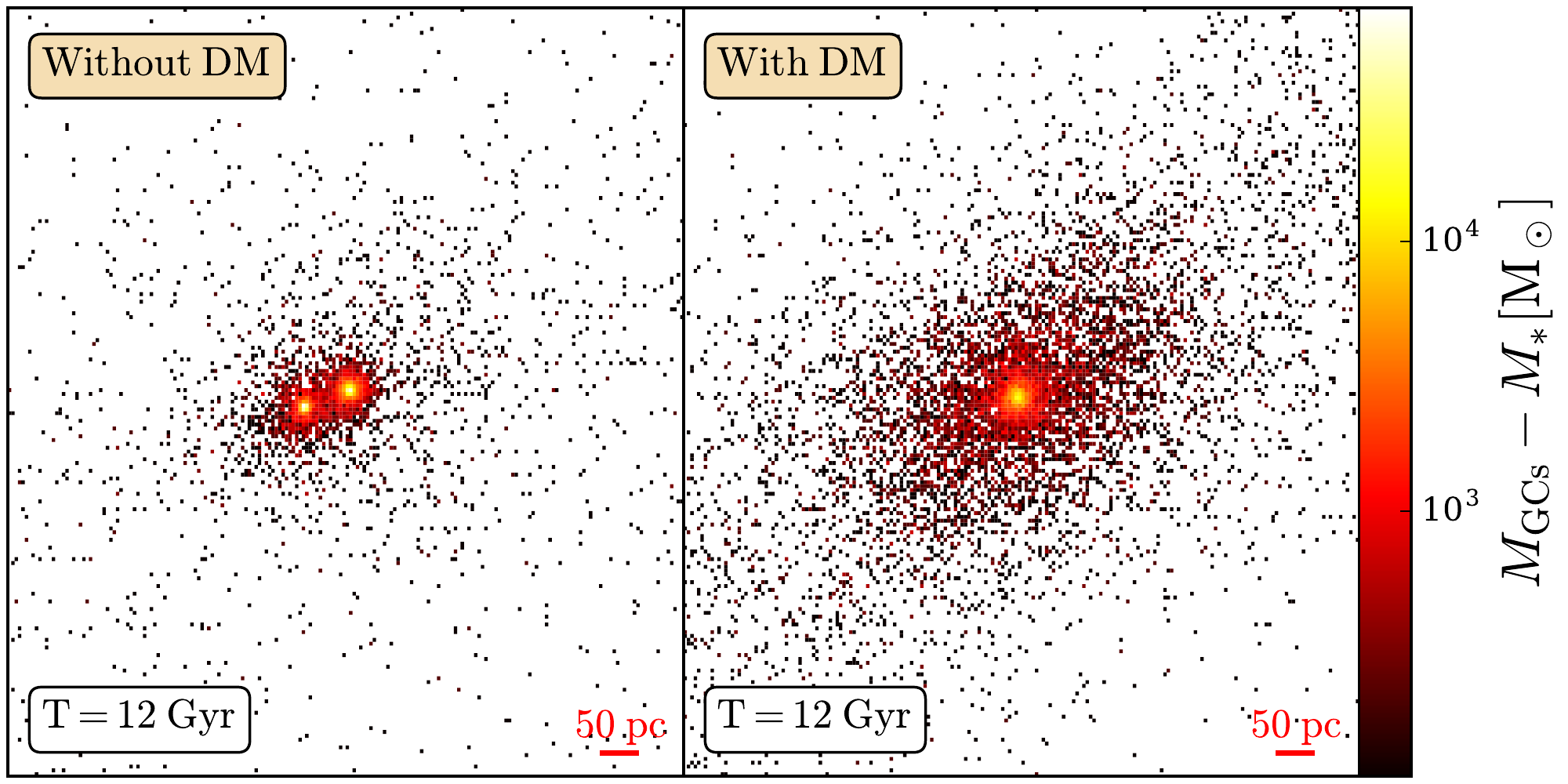}
\caption{{\it Formation of a nuclear star cluster (NSC):} Mass difference between stars of the 5 GCs without ({\it left panel}) and with ({\it right panel}) a DM minihalo and Fornax stars at $T=12$ Gyr. We represented only bins with a size of 5 pc, which have a positive mass difference. The latter allows us to highlight a stellar overdensity induced by GCs at the Fornax centre. This mass difference exhibits the presence of a double nucleus for stellar GCs and single nucleus for GCs with DM minihalo. Infalling GCs, which are not completely tidally stripped, will contribute to the formation of a NSC. As Fornax exhibits the absence of a nuclear star cluster at the centre, the formation of a NSC ruled out the early accretion scenario for Fornax GCs.}
\label{f1}
\end{figure}

\begin{figure}
\centering
\includegraphics[width=0.47\textwidth]{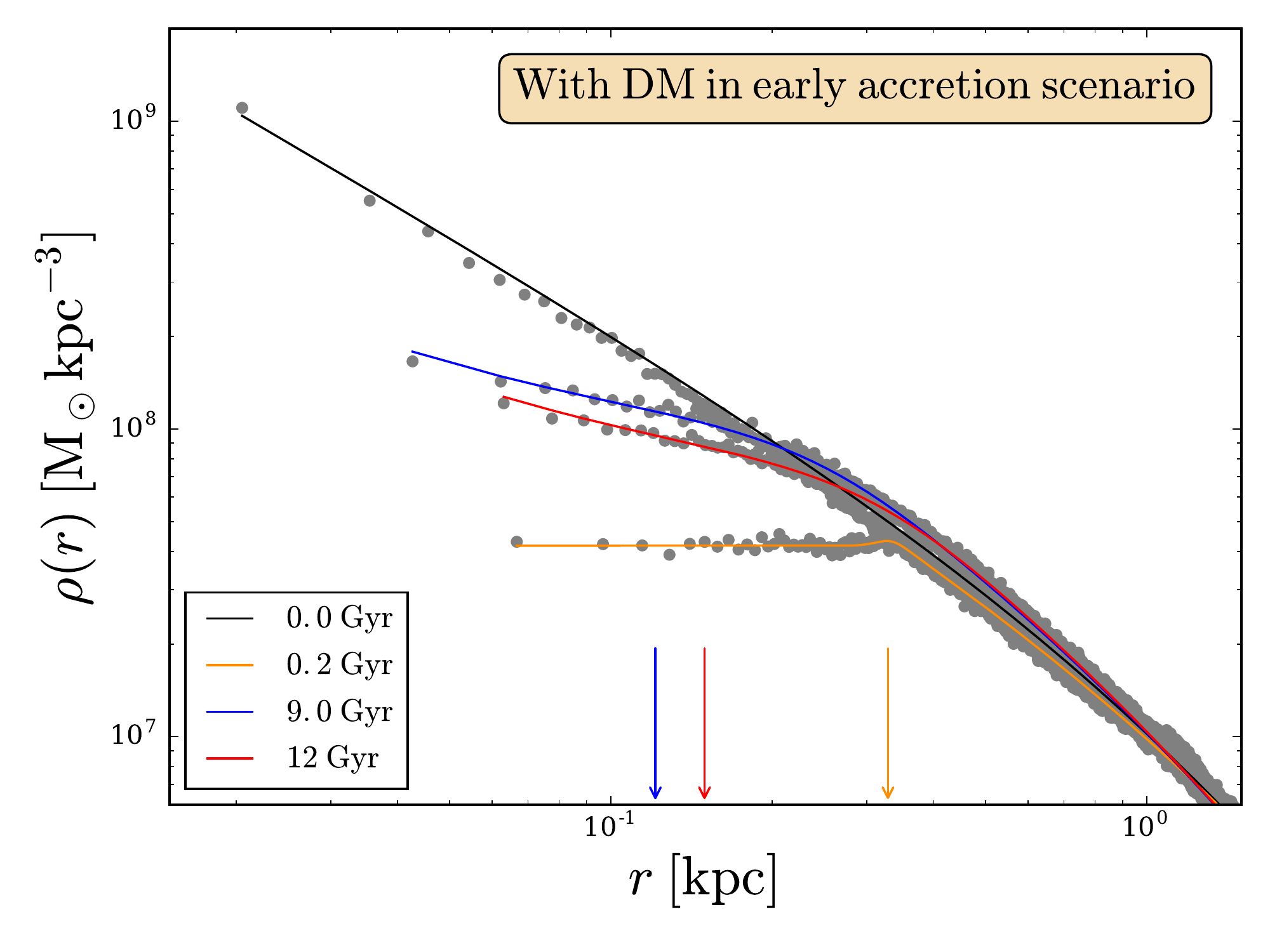}
\caption{{\it Heating of the DM central region:} DM density profile of Fornax at different times with its corresponding core radius (marked by arrows). We represent only cored profiles due to the heating by GCs with a DM minihalo in the early accretion scenario. Initially, the Fornax DM halo assumes a NFW profile. We consider DM particles from both Fornax halo and GCs to determine the DM density profile of Fornax. The fitting function described by Equation~\ref{eq1} reproduces the simulated density structures and captured the rapid transition from the cusp to the core. We set Poissonian errors for fitting weights. The DM distribution of Fornax halo is divided in bins of groups composed of $N_{\rm g}=1024$ particles.}
\label{fgnew}
\end{figure}

\begin{figure}
\centering
\includegraphics[width=0.47\textwidth]{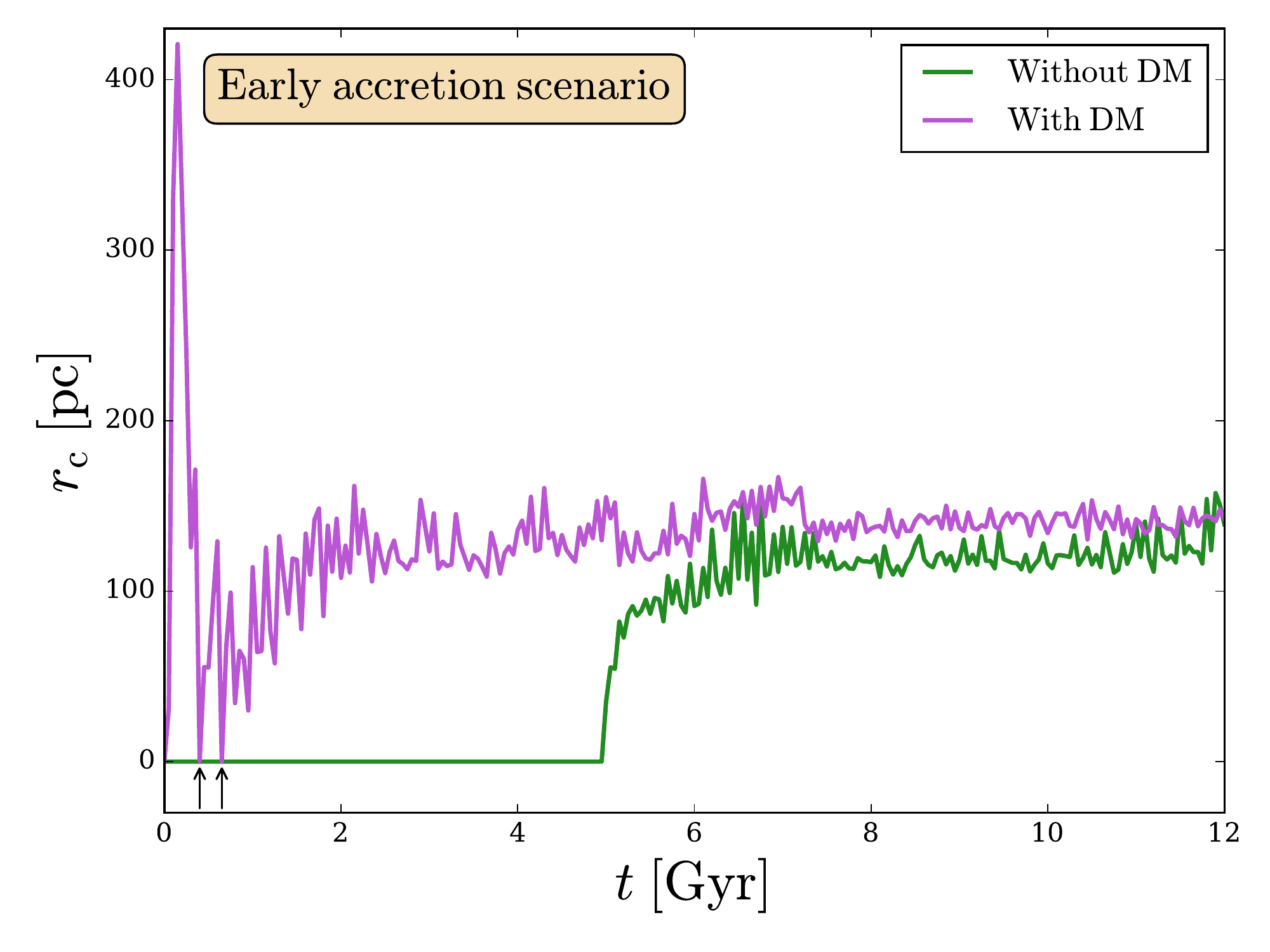}
\caption{{\it Fornax DM core in the early accretion scenario:} Fitted core radius $r_{\mathrm{c}}$ of the DM cored halo (see Equation~\ref{eq1}) induced by crossings of GCs with (in purple) and without (in green) a DM minihalo as a function of time. $r_{\mathrm{c}}\neq 0$ ($r_{\mathrm{c}}=0$) means that there is a (no) cusp-to-core transition for the Fornax DM halo. The absence of a core within the first 5 Gyr states that stellar GCs in orbit cannot generate DM cores due to their low mass impact. However, once they are spiralling into the centre of the galaxy, they induce a cusp-to-core transition. Contrary to stellar GCs, dynamical heating of the DM field from DM minihalo crossings drives the core formation. The core sizes depend on the frequency of the crossings. Besides, between crossings, the halo can re-form the cuspy halo owing to the new orbits of DM particles initially at the Fornax centre as they gained energy from the GCs. We observed these reverse transitions (marked by black arrows) of the Fornax DM halo.}
\label{fgr6}
\end{figure}

First, we assume that GCs with and without a DM minihalo were accreted 10-12 Gyr ago by Fornax. Fig.~\ref{fgr2a} depicts the orbital decay of five GCs with and without a DM minihalo over 12 Gyr in this scenario. Initially, they have the same most prevalent positions and velocities at $z=3$ from Illustris TNG-100 cosmological simulation (see objects $E_i$ in Table~\ref{tab1}). Our initial conditions entail an accretion process by the Fornax galaxy with eccentric orbits for all the GCs. The orbital period is higher for the stellar GCs than for GCs with a DM minihalo. In Fig.~\ref{fgr2a}, orbital radii confirm that DM minihalos accelerate the infall of the five GCs. The stellar mass loss of all the GCs over 12 Gyr is presented in Fig.~\ref{fgrb}. GCs with DM minihalos lose more stars compared to stellar GCs, because minihalos induce  major tidal stripping. Indeed, stripping by the Fornax tidal field is more efficient at small radii. We also observe that three GCs with DM minihalos are completely tidally stripped within 8 Gyr.

Fornax is dominated by metal-rich stars whereas GCs are dominated by metal-poor stars. The metal-poor stellar mass in Fornax was estimated to be about (44.9 $\pm$ 5.3)$\times10^{5}$ $M_{\sun}$ \citep{2016A&A...590A..35D}. This large quantity of metal-poor stars in Fornax could suggest that each of these four surviving metal-poor GCs has likely lost several times $10^{5}$ $M_{\sun}$ due to dynamical processes. GC4 is excluded from this estimate, because this cluster is possibly more metal-rich than the other clusters. Another explanation is that GCs have sunk to the galaxy centre and were destroyed such as our three GCs with DM minihalos. Hence, these metal-poor stars could correspond to relics of destroyed GCs with DM minihalos. This could be also possible for stellar GCs, but this scenario within 12 Gyr is more reliable for DM minihalos as the latter accelerate the tidal stripping (see in Fig.~\ref{fgrb}). Fornax observations reveal that Fornax has no NSC at its centre. In order to verify that no NSC was formed in our scenario, Fig.~\ref{f1} illustrates the mass difference between stars of the 5 GCs, without and with a DM minihalo, and Fornax stars at $T=12$ Gyr. We represent only bins with a size of 5 pc, which have a positive mass difference. The latter allows us to highlight an overdensity for stars. In other words, we want to clearly separate a NSC from the stellar component of the galaxy. Fig.~\ref{f1} showed the presence for a double nucleus for stellar GCs and single nucleus for GCs with DM minihalo formed by infalling GCs, which are not completely destroyed by the Fornax tidal field. Thus, we rule out the early accretion scenario for Fornax GCs due to the infall of the five GCs and the formation of a NSC. We point out that stars of GCs with a DM minihalo, which are completely tidally stripped, will not contribute to the formation of a nuclear star cluster at the centre.

Our GPU simulations allow us to also study the impact of GCs with or without a DM minihalo on the evolution of the DM density profile of Fornax. We observe the formation of cores in the cold dark matter halo of Fornax. As all GCs have eccentric orbits, they can perturb the DM halo of Fornax by their multiple crossings near the Fornax centre (see Fig.~\ref{fgr2a}). Initially, the Fornax DM halo assumes a NFW profile. We consider DM particles from both Fornax halo and GCs to determine the DM density profile of Fornax over the time. In order to determine precisely if there is  core formation in the Fornax halo, we did a fit for the DM profile. As shown in Fig.~\ref{fgnew}, we found that all our profiles are well fitted by the following five-parameter formula:

\begin{equation}
    \rho(r) = \rho_{\mathrm{c}} W(r) + [1 - W(r)]\rho_{\mathrm{NFW}}(r),
\label{eq1}
\end{equation}
where $\rho_{\mathrm{c}}$ is the core constant density and $W(r)$ is defined as
\begin{equation}
2W(r)=1-\erf\left(\frac{r-r_{\mathrm{c}}}{2\delta}\right),
\end{equation}
where $r_{\mathrm{c}}$ is the core radius and $\delta$ is a parameter to control the sharpness of the transition from the cusp to the core and the converse. It reproduces the simulated density structures and captured the rapid transition from the cusp to the core (see Fig.~\ref{fgnew}). We set Poissonian errors for fitting weights. The DM distribution of Fornax halo is divided in bins of groups composed of $N_{\rm g}=1024$ particles. The smallest core size in the simulations is of the order of 10 pc, which corresponds to our spatial resolution. Thus, we assume that a cusp-to-core transition occurred when the core size $r_{\mathrm{c}}$ becomes greater than our spatial resolution. As the value of the core size depends entirely on the fitted DM profile, we did not use this radius as a constraint on the DM halo core size of Fornax. Fig.~\ref{fgr6} describes the evolution of the fitted core radius $r_{\mathrm{c}}$ of the Fornax halo over 12 Gyr due to heating by GCs with and without a DM minihalo. We consider the core radii $r_{\mathrm{c}}$ to determine whether a transition appears in our simulation. Non-zero core sizes $r_{\mathrm{c}}$ means that a cusp-to-core transition occurred for the Fornax DM halo. The absence of cores ($r_{\mathrm{c}}=0$) means that DM halo is cuspy and is still described by a NFW profile. For stellar GCs, the absence of core shows that these $10^{6}$ $M_{\sun}$ orbiting objects can't generate DM cores. The energy transfer is not sufficient to perturb the DM distribution at the centre of the galaxy due to their low mass. However, once they are spiralling into the centre of the galaxy, they induce a cusp-to-core transtion. Then, we observe a DM core of about 100-150 pc for the Fornax halo. Contrary to stellar GCs, dynamical heating of the DM field from DM minihalo crossings drives core formation. These core sizes are between 150 and 400 pc, and depend on the frequency of the crossings. Besides, between crossings, the halo can re-form the cuspy halo owing to the new orbits of DM particles initially at the Fornax centre as they gained energy from the GCs. Fig.~\ref{fgr6} shows cusp-to-core transitions and the reverse transitions (marked by black arrows) of the Fornax DM halo. In this early accretion scenario, GCs with and without a DM minihalo cannot explain both DM core formation and timing problem. Despite the core formation in cold dark matter, they are ruled out by the GC infall and consequent NSC formation.   

\subsection{Recent accretion}

\begin{figure}
\centering
\includegraphics[width=0.47\textwidth]{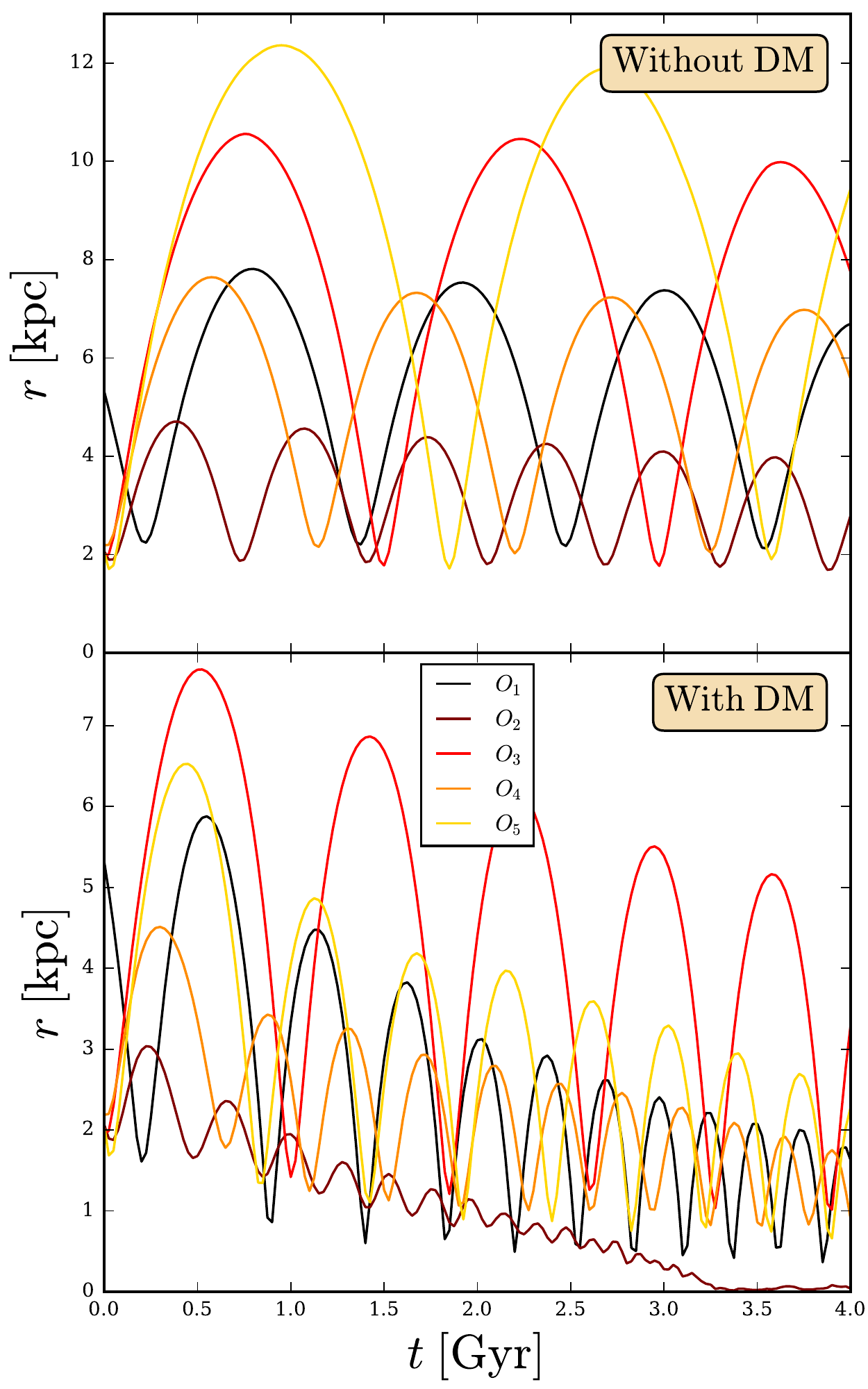}
\caption{{\it DM minihalos stay in orbit:} Orbital decay of the 5 GCs without ({\it upper panel}) and with ({\it lower panel}) a $2\times10^7$ M$_{\sun}$ DM minihalo over 4 Gyr. These radii correspond to the distances between the mass centre of each GC and Fornax. Initially, $10^6$ M$_{\sun}$ GCs with and without DM minihalo have the same most prevalent positions and velocities at $z=0.36$ from Illustris TNG-100 cosmological simulation (see objects $O_i$ in Table~\ref{tab1}). Our initial conditions entail an accretion process by Fornax galaxy only for GCs with a DM minihalo. Stellar GCs experienced stable eccentric orbits beyond 2 kpc.}
\label{fgr2b}
\end{figure}

\begin{table}
\begin{center}
\label{tab:landscape}
\begin{tabular}{cccccccccc}
 \hline
    Object & $M^{(a)}_{\mathrm{obs}}$ & $D_{\mathrm{obs}}^{(b)}$ & & $r$ & $M$ \\
     & [$10^{5}$ $M_{\sun}$] & [kpc]&  & [kpc]& [$10^{5}$ $M_{\sun}$]\\
    \hline
    GC1 & 0.42 $\pm$ 0.10 & 1.6 & $O_3$ & 5.38 & 9.98 \\
    GC2 & 1.54 $\pm$ 0.28 & 1.05 & $O_5$ & 3.26 & 9.91 \\
    GC3 & 4.98 $\pm$ 0.84 & 0.43 & $O_1$ & 2.33 & 8.57 \\
    GC4 & 0.76 $\pm$ 0.15 & 0.24 & $O_2$ & 0.28 & 6.83 \\
    GC5 & 1.86 $\pm$ 0.24 & 1.43 & $O_4$ & 1.64 & 9.67 \\
    \hline 
\end{tabular}
\caption{Comparison with GC observations. The final radii and masses of GCs embedded in DM minihalos at 3 Gyr. We found GC candidates $O_i$ compatible with the observed projected distances $D_{\mathrm{obs}}$. However, we found higher masses for simulated GCs than for those observed. We note that we could set lower stellar mass to the initial GCs in order to reproduce the observed masses. DM minihalo of GCs solve the Fornax timing problem if GCs were accreted less than 3 Gyr ago. References: (a) \protect\cite{2016A&A...590A..35D}, (b) \protect\cite{2003MNRAS.340..175M}.}
\label{tab2}
\end{center}
\end{table}

\begin{figure}
\centering
\includegraphics[width=0.47\textwidth]{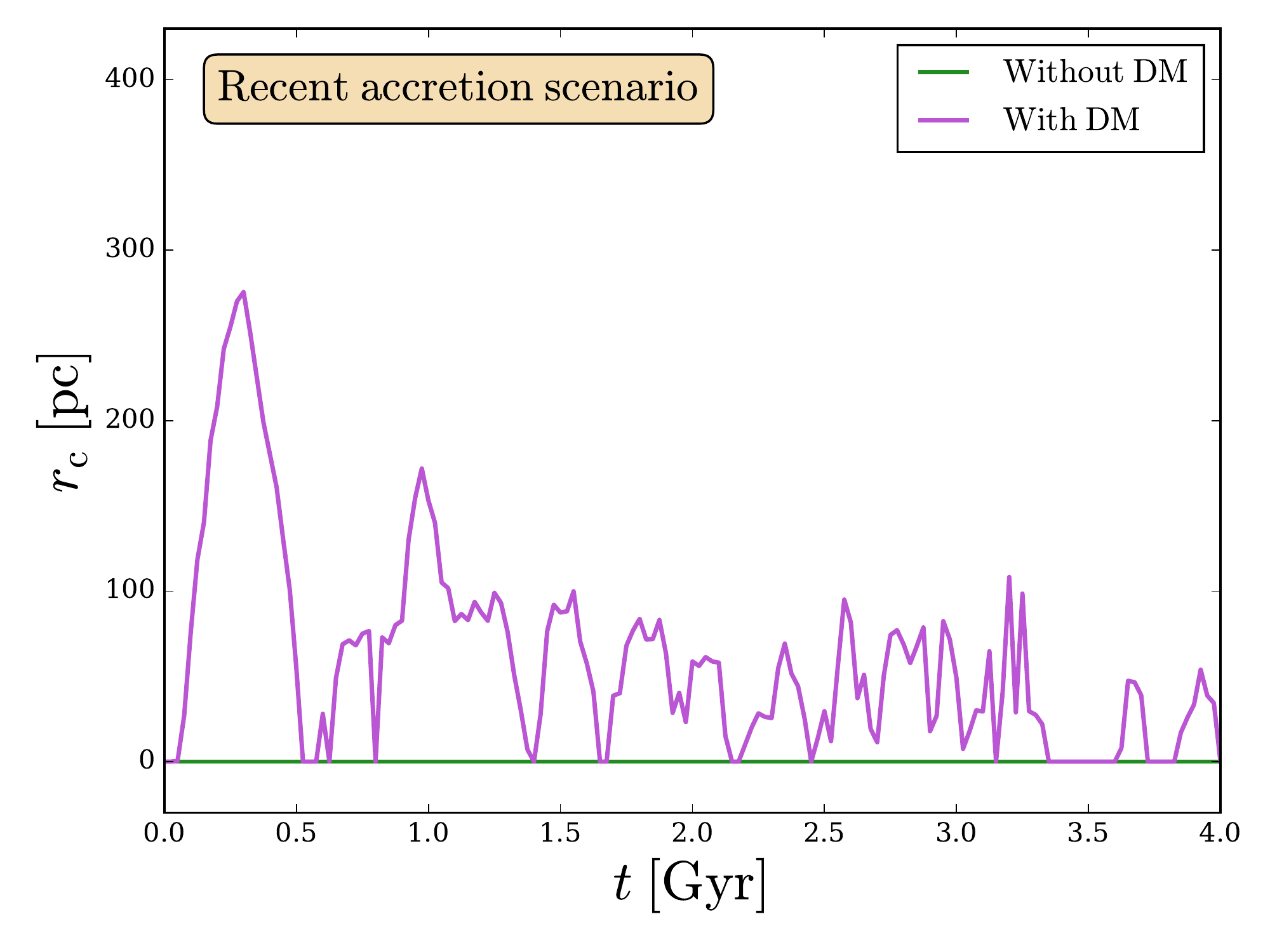}
\caption{{\it Fornax DM core in the recent accretion scenario:} Fitted core radius $r_{\mathrm{c}}$ of the DM cored halo induced by crossings of GCs with (in purple) and without (in green) a DM minihalo as a function of time. $r_{\mathrm{c}}\neq 0$ ($r_{\mathrm{c}}=0$) means that there is a (no) cusp-to-core transition for the Fornax DM halo. In this scenario, the complete absence of core confirms again that stellar GCs on orbit cannot generate DM cores due their low mass impact. Contrary to stellar GCs, dynamical heating of the DM field from DM minihalo crossings entails the core formation.}
\label{fgr6b}
\end{figure}

\begin{figure}
\centering
\includegraphics[width=0.47\textwidth]{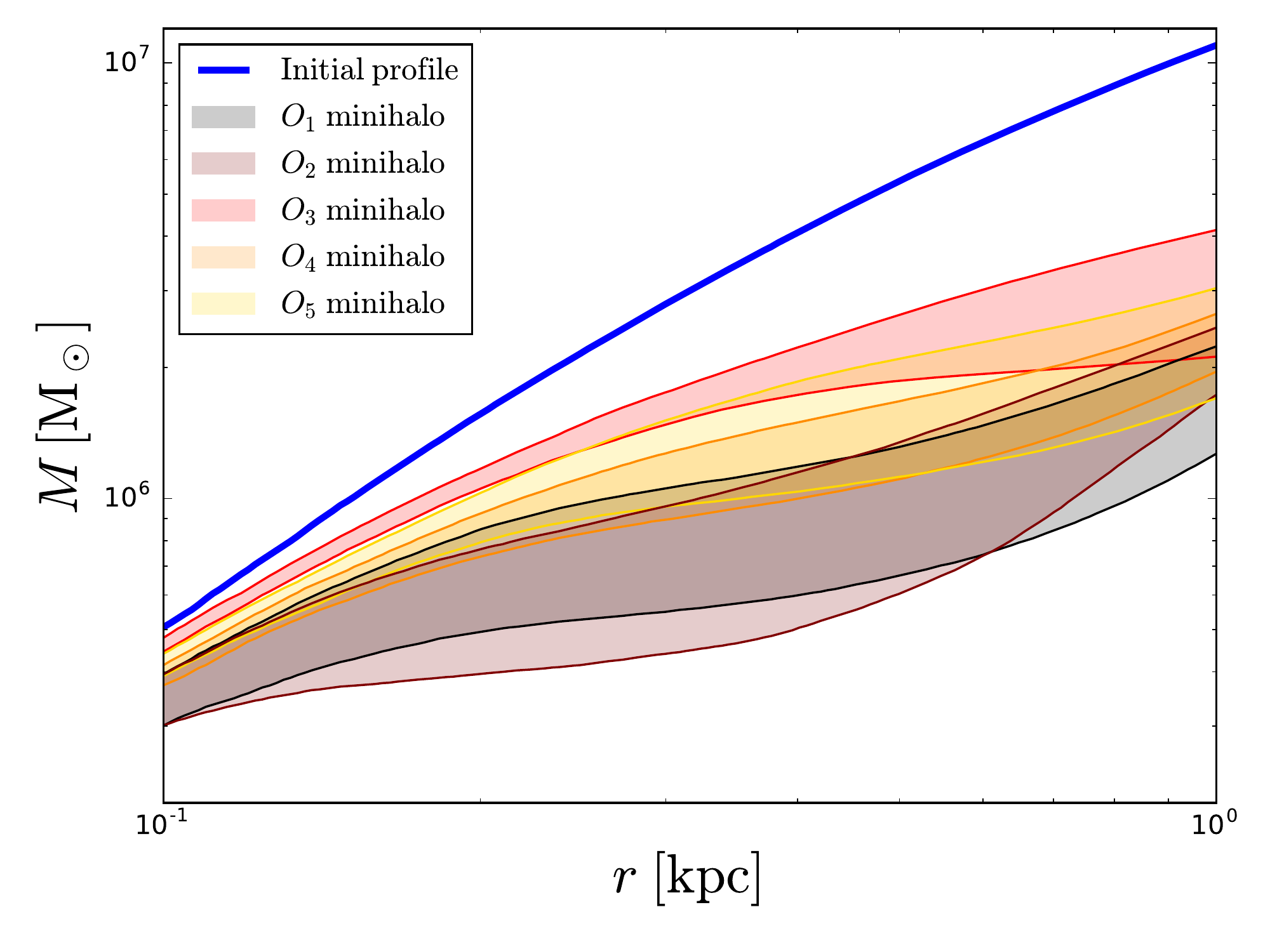}
\caption{{\it DM minihalo remnants:} Masses of remnant DM minihalos centered on the stellar component of $O_i$ GCs between 2 and 3 Gyr. The blue dashed line represents the initial distribution of $2 \times 10^{7}$ M$_{\sun}$ minihalo centered on its stellar component. It is shown that all the DM minihalos have been tidally stripped by the tidal field of Fornax. We have found that GCs are embedded in DM minihalos less massive than $10^{7}$ $M_{\sun}$ inside the central 500 pc after 2-3 Gyr, which is in agreement with the observed prediction on a MW GC, NGC 2419 \citep{2009MNRAS.396.2051B,2013MNRAS.428.3648I}}
\label{fgr3}
\end{figure}

In this section, we assume that GCs with and without a DM minihalo were accreted 4 Gyr ago by Fornax. Fig.~\ref{fgr2b} describes the orbital decay of GCs with and without a DM minihalo over 4 Gyr in this scenario. Initially, they have the same most prevalent positions and velocities at z=0.36 from Illustris TNG-100 cosmological simulation (see objects $O_i$ in Table~\ref{tab1}). According to the initial orbital radii, stellar GCs ({\it upper panel}) stay in orbit beyond 2 kpc from the centre, whereas GCs with a DM minihalo ({\it lower panel}) are accreted by Fornax and are falling towards the Fornax centre. But, as all the 5 GCs with a DM minihalo are still orbiting before 3 Gyr, they cannot form a NSC in accordance with observations. Table~\ref{tab2} depicts orbital radii and masses of GCs embedded in a DM minihalo at 3 Gyr. For each observed Fornax GC, we propose a GC with a DM minihalo as a candidate in order to reproduce the spatial distribution of observed GCs if GCs were accreted less than 3 Gyr ago. Their orbital radii are higher than the projected distances $D_{\mathrm{obs}}$, which are the minimum distances between observed GCs and the Fornax centre. Concerning the GC observed mass, we found a higher mass from the simulations for each GC with a DM minihalo. As the GC dynamical evolution is entirely determined by the DM minihalo, we could easily set lower stellar mass limits to the initial GCs in order to reproduce the observed masses. Reducing the stellar mass of GCs could contribute to their survival. Similarly to GCs with a DM minihalo, the spatial distribution of stellar GCs over 4 Gyr is also compatible with observations of Fornax GCs. Thus, both GC types can resolve the Fornax timing problem in this recent accretion scenario. 

Nevertheless, Fig.~\ref{fgr6b} highlights that there is no cusp-to-core transition for stellar GCs in this scenario. As they stay on orbit beyond 2 kpc, they cannot transfer energy to DM particles at the centre of the galaxy. Moreover, we established previously that their crossings cannot perturb the DM halo due to their low mass compared to DM minihalos. With stellar GCs, the DM profile does not change over time as shown by \cite{2019MNRAS.485.2546B}. Concerning GCs with a DM minihalos, we observe a cusp-to-core transition induced by their crossings. More precisely, there are forward and reverse transitions from the cusp to the core. However, most of the time, Fornax is expected to have a core due to DM minihalo crossings according to our simulation results. In addition, in the recent accretion scenario, we show that only GCs with a DM minihalo can explain both the DM core formation and timing problem in Fornax. 

As the recent accretion scenario with GCs with a DM minihalo is compatible with Fornax observations, especially for the DM core formation and GC spatial distribution, we are interested by the mass loss of DM minihalos. Fig.~\ref{fgr3} illustrates the mass of remnant DM minihalos centered on the stellar component of the five GCs $O_i$ between 2 and 3 Gyr. It is shown that all the DM minihalos have been tidally stripped by the tidal field of Fornax. Even if GCs are not proven to have a significant amount of DM, this does not preclude them from having been formed originally inside a DM minihalo. Indeed, we showed that our GCs lost a large fraction of their DM minihalos. We have found that GCs are embedded in DM minihalos less massive than $10^{7}$ M$_{\sun}$ inside the central 500 pc after 2-3 Gyr, which is also in agreement with the observed prediction on a MW GC, NGC 2419, based on the observed velocity dispersion \citep{2009MNRAS.396.2051B}. We found also  good agreement with the prediction of \cite{2013MNRAS.428.3648I}, who established that the virial mass of the minihalo of NGC2419 cannot exceed $\sim$ $4 \times 10^{6}$ M$_{\sun}$.

\subsection{Enhancement of the core formation}

\begin{figure}
\centering
\includegraphics[width=0.47\textwidth]{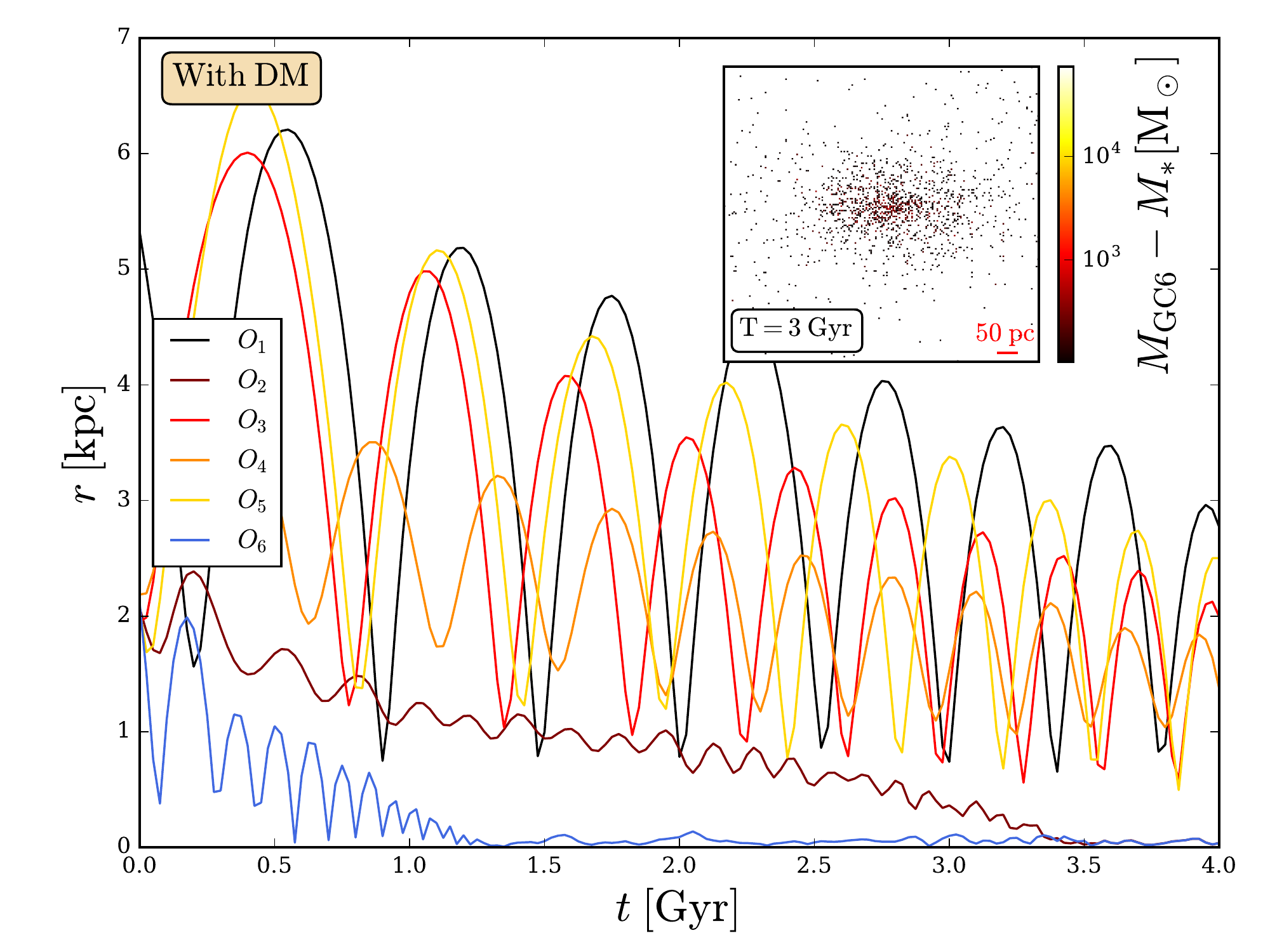}
\caption{{\it Infall without forming a NSC:} Orbital decay of the 6 GCs with a $2\times10^7$ M$_{\sun}$ DM minihalo over 4 Gyr. These radii correspond to the distances between mass centre of each GC and Fornax. The first five GCs have the same initial conditions as in Fig.~\ref{fgr2b}. Stellar distribution of the mass difference between the $O_6$ GC with a DM minihalo and Fornax stellar component ({\it inset}) highlights that there is no NSC at the centre of the galaxy despite the rapid infall of this additional GC with a DM minihalo. Finally, we establish that Fornax could have more than five 5 GCs. The extra GCs have fallen to the Fornax centre and were destroyed, which is compatible with the large quantity of the metal-poor stars found in Fornax. The absence of NSC on this timescale is only possible because DM minihalos accelerate the tidal stripping.}
\label{fgr2c}
\end{figure}

\begin{figure}
\centering
\includegraphics[width=0.47\textwidth]{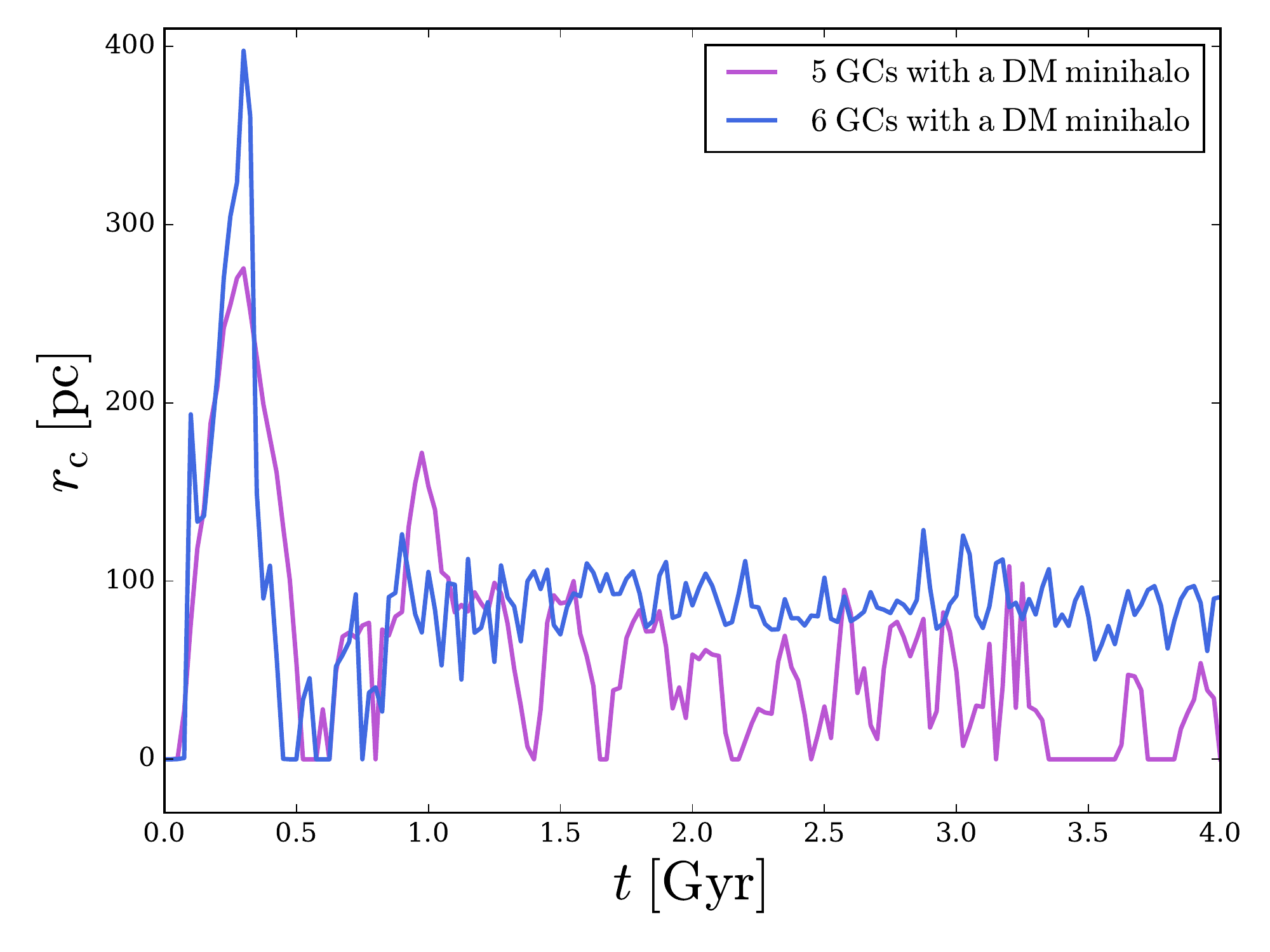}
\caption{{\it Enhancement of the core formation:} Fitted core radius $r_{\mathrm{c}}$ of the DM cored halo induced by crossings of the five $O_i$ GCs with a DM minihalo (in purple) plus one additional GC with a DM minihalo (in blue) as a function of time. $r_{\mathrm{c}}\neq 0$ ($r_{\mathrm{c}}=0$) means that there is a (no) cusp-to-core transition for the Fornax DM halo. In this scenario, the rapid infall of $O_6$ GC with a DM minihalo enhances the core formation due to crossings of the five GCs with DM minihalos.}
\label{fgr6c}
\end{figure}

For both GCs with and without a DM minihalo, we noticed previously that spiralling GCs at the centre of the galaxy enhance the core formation (see in Fig.~\ref{fgr2a} and~\ref{fgr6}). However, the infall of GCs can entail the formation of a NSC at the centre of the galaxy. For NSC formation, GCs at the centre have to be completely tidally stripped. Thus, the GC tidal stripping needs to be accelerated as in the case of the DM minihalo (see in Fig.~\ref{fgrb}). Adding an infalling GC with a DM minhalo is also motivated by the fact that Fornax has a large quantity of metal-poor stars, which could correspond to relics of destroyed GCs with  DM minihalos. In addition, according to its stellar kinematics, Fornax is expected to have a large core \citep{2011MNRAS.411.2118A,2011ApJ...742...20W,2018MNRAS.480..927P,2019MNRAS.482.5241K,2019MNRAS.484.1401R}. Adding a falling GC with a DM minihalo, which is going to be completely disrupted, could contribute to the formation of a larger DM core. We test this hypothesis by running a simulation with the five $O_i$ GCs with a DM minihalo and one additional GC with a DM minihalo. We want to improve the degree of core formation in the recent accretion scenario. Fig.~\ref{fgr2c} describes the orbital decay of the 6 GCs with a $2\times10^7$ M$_{\sun}$ DM minihalo over 4 Gyr. The first five GCs have the same initial conditions as in Fig.~\ref{fgr2b}. We notice that $O_6$ GC with a DM minihalo spirals to the centre after 1 Gyr. Stellar distribution of the mass difference between the $O_6$ GC with a DM minihalo and Fornax stellar component ({\it upper subplot}) highlights that there is no NSC at the centre of the galaxy despite the rapid infall of this additional GC with a DM minihalo. As expected, $O_6$ GC is completely destroyed after 3 Gyr (see inset in Fig.~\ref{fgr2c}). Finally, we establish that Fornax could have more than five 5 GCs. More than one extra GC could have fallen to the Fornax centre a long time ago and could have been destroyed, which is compatible with the large quantity of the metal-poor stars found in Fornax. The absence of NSC on our timescale is only possible because DM minihalos accelerate the tidal stripping. Fig.~\ref{fgr6c} compares the fitted core radius as a function of time for the five $O_i$ GCs with a DM minihalo (in purple) and these five GCs plus one additional GC with a DM minihalo (in blue). We show that the rapid infall of $O_6$ GC with a DM minihalo enhances the core formation due to crossings of the five GCs with a DM minihalo. Thus, if Fornax had more than five 5 GCs, they could contribute to form a larger core.

As the value of the core size $r_{\mathrm{c}}$ depends entirely on the fitted DM profile, we can't use this radius as a constraint on the DM halo core size of Fornax. However, the DM density at 150 pc of Fornax was estimated to be around $10^8$ M$_{\sun}$.$\mathrm{kpc^{-3}}$  for a virial mass of $\sim 2\times10^{10}$ M$_{\sun}$ by using stellar kinematics \citep{2019MNRAS.484.1401R}. Assuming a virial mass of $10^9$ M$_{\sun}$, we found a similar density at this specific radius in our simulation with 6 GCs embedded in a DM minihalo. As core formation is entirely driven by energy transfers via dynamical friction, it will be harder to induce a core with this density at 150 pc for a $10^{10}$ M$_{\sun}$ halo with only five globular clusters, because the center of the higher mass halo is less dense. Thus, dynamical friction will be less efficient in these halos. Nevertheless, DM substructures will have the same behaviour as our GCs with a DM minihalo. As the GC dynamical evolution is entirely determined by the DM minihalos, these substructures could have also interacted with Fornax and induced a cusp-to-core transition. In addition to GCs, subhalos could be responsible for the large core formation in Fornax. The dynamics of DM subtructures in dwarf galaxies is investigated in Boldrini et al. (in prep.).

\section{Conclusion}

In this work, we have studied the motion of GCs embedded in DM minihalos inside the cold DM halo of Fornax in order to deal with the timing problem. We have considered an early and a recent accretion scenario of GCs by Fornax with the most prevalent initial conditions taken from Illustris TNG-100 cosmological simulations. Using a fully GPU $N$-body code, we propose  a new mechanism to resolve the cusp-core problem. First, the infall of GCs and the formation of a nuclear star cluster rules out the early accretion scenario for GCs with and without a DM minihalo. However, we showed that GC crossings near the Fornax centre induce a cusp-to-core transition of the DM halo. Secondly, we demonstrated that DM minihalos, as a new component of GCs, resolve both the timing and cusp-core problems in Fornax if the five GCs were accreted recently, less than 3 Gyr ago, by Fornax. Under these assumptions, the infall of these GCs does not occur and no star cluster forms in the centre of Fornax in accordance with observations. Crossings of GCs with a DM minihalo near the Fornax centre induce a cusp-to-core transition of the DM halo and hence resolve the cusp-core problem in this dwarf galaxy. The DM core size depends strongly on the frequency of GC crossings. We subsequently highlighted that an infalling GC with a DM minihalo enhances core formation without forming a NSC at the Fornax centre. Moreover, we are in good agreement with the constraints on the DM mass of GCs as our clusters lost a large fraction of their DM minihalos. All of these aspects provide strong evidence for the existence of DM halos in GCs.

Our simulations clearly show  that between  central passages, the DM halo can regenerate its cusp. \cite{2019MNRAS.484.1401R} found that dwarf galaxies can be separated into two distinct classes, those with cold DM cusps and DM cores. Fornax favours a DM core, whereas Carina, Sextant and Draco favour a DM cusp. The transient phenomenon that we have found could explain this diversity of DM halo profiles. GCs embedded in  DM minihalos, which are eventually completely stripped, could have induced such cup-core  transitions by past infall. However, we regard this as unlikely since GCs do not appear to be ubiquitous in dwarf galaxies.

\section*{Acknowledgments}
We thank our anonymous referee for  helpful comments and suggestions that improved our work. We thank also Miki Yohei for providing the non-public $N$-body code, \textsc{gothic}. We would like to thank Apolline Guillot, Dante Von Einzbern, Colandrea Guillot and George Bool for their constructive suggestions to improve the manuscript. 

\section*{Appendix}

\begin{figure}
\centering
\includegraphics[width=0.47\textwidth]{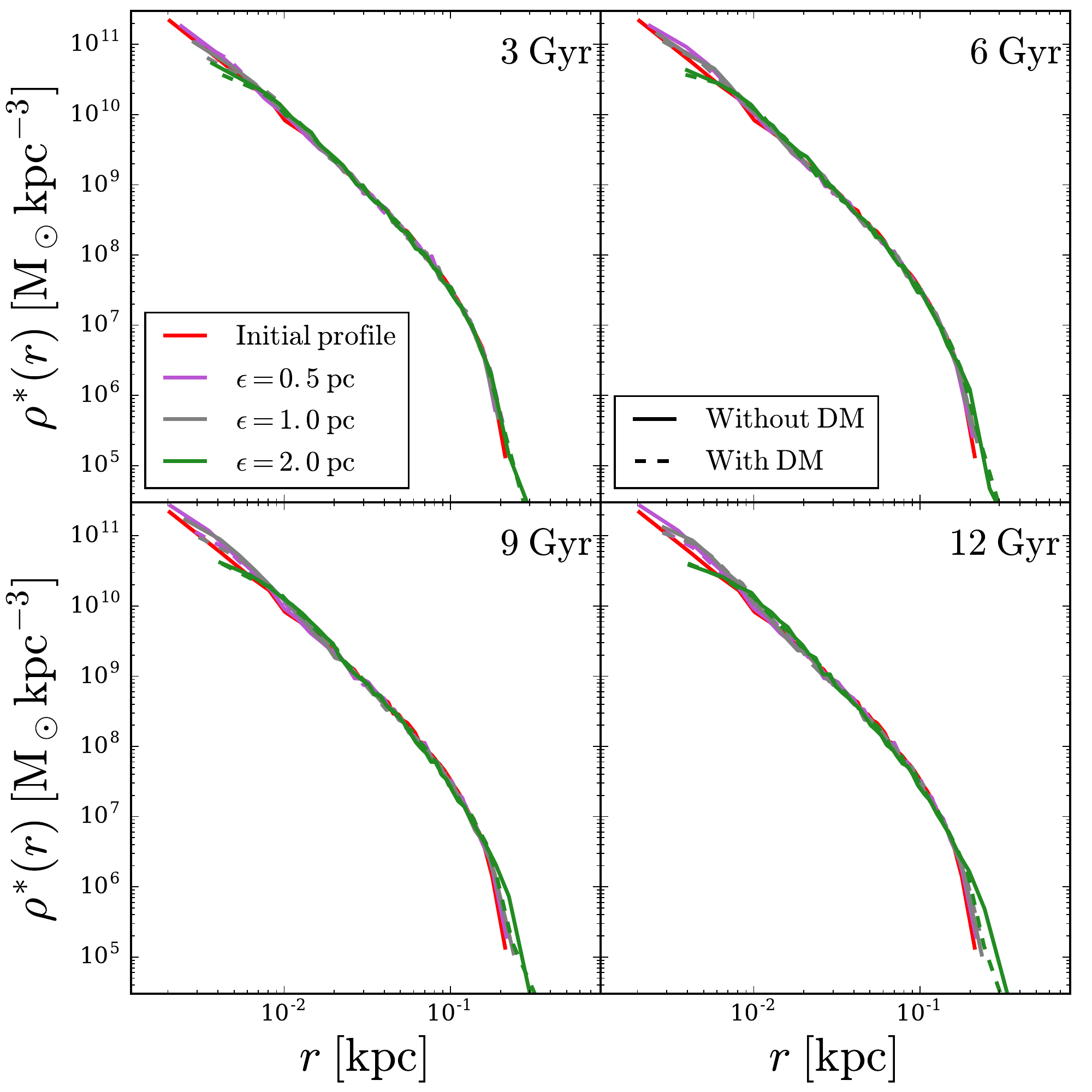}
\caption{{\it Impact of softening on GC stellar density profiles:} Stellar density profile of an isolated GC with and without a DM minihalo for different times and three different softening lengths. Initially, GC stars assume a King profile with a King radius $r_{\rm k} = 1$ pc. The stellar distribution is divided in bins of groups composed of $N_{\rm g}=216$ particles. Our mass resolution does not allow us to resolve the 1 pc core radius of GCs. We noticed that the dynamics of GCs is subject to numerical effects for $\epsilon=2$ pc. Our convergence test states that the stellar density profiles for $\epsilon=0.5$ and 1 pc are nearly identical.}
\label{app21}
\end{figure}

\begin{figure}
\centering
\includegraphics[width=0.47\textwidth]{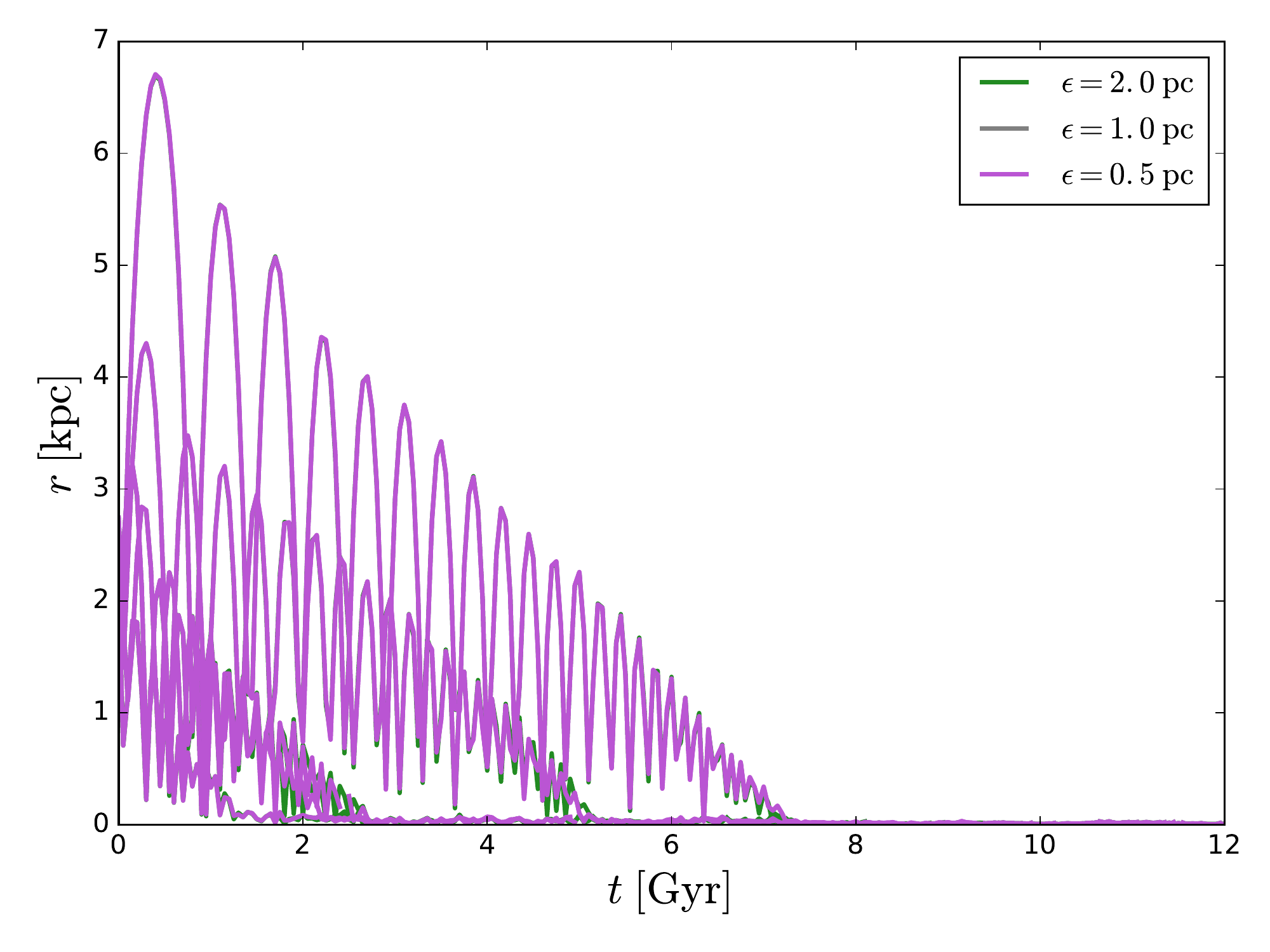}
\caption{{\it Impact of softening on orbital decay:} Orbital radius of the five GCs with a DM minihalo in an early accretion process with eccentric orbits (see objects $E_i$ in Table~\ref{tab1}) as a function of time for three different softening lengths. The orbital decays of GCs are nearly identical for all the softening lengths. As the stellar density profiles for $\epsilon=0.5$ and 1 pc are very similar (see Fig.~\ref{app21}), the orbital radius behaves the same for these softenings.}
\label{app1}
\end{figure}

\begin{figure}
\centering
\includegraphics[width=0.47\textwidth]{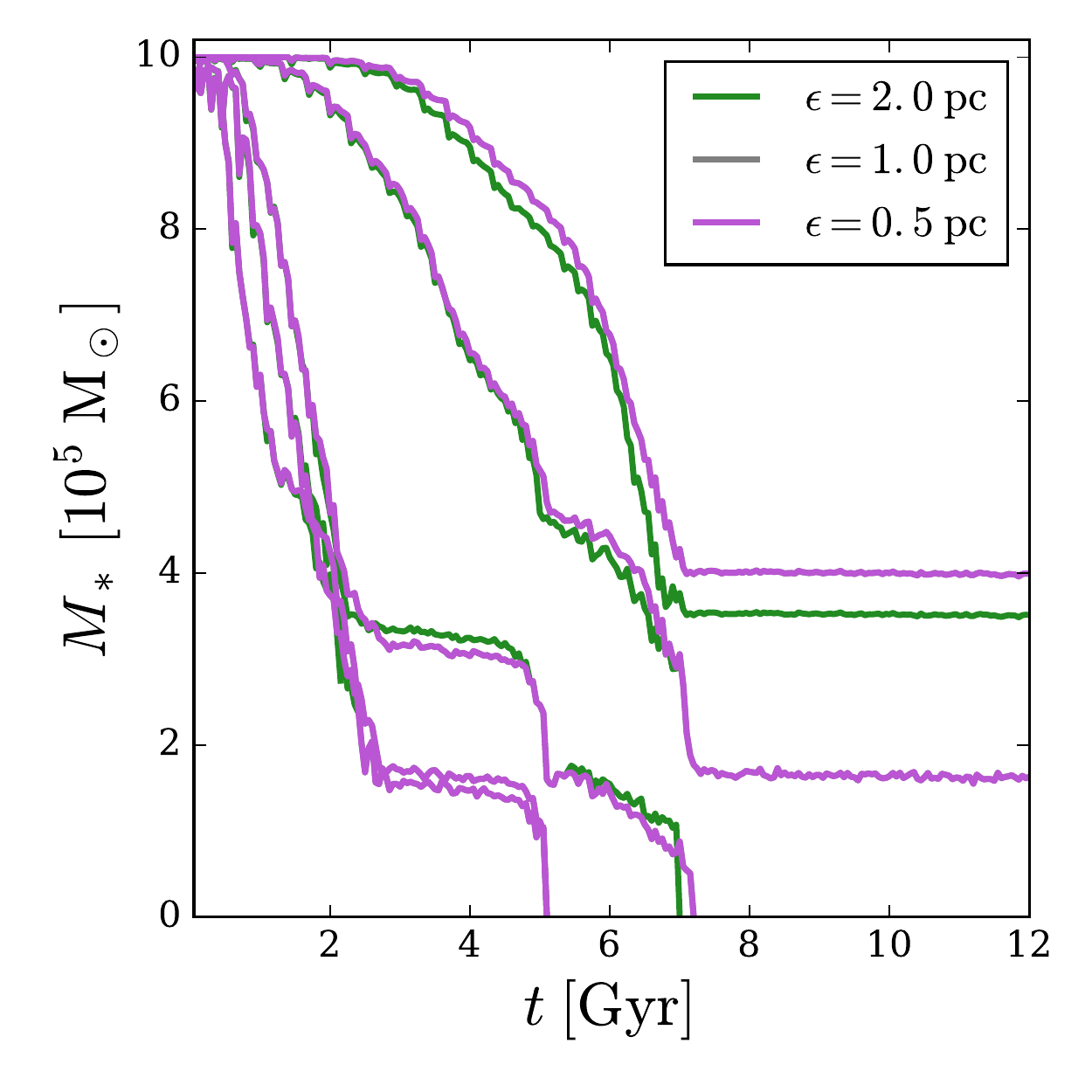}
\caption{{\it Impact of softening on mass loss:} Evolution of the mass loss of the stellar component embedded in a DM minihalo over 12 Gyr for three different softening lengths. Our simulations are well converged for $\epsilon=0.5$ and 1 pc. As the stellar density profiles for $\epsilon=0.5$ and 1 pc are nearly identical (see Fig.~\ref{app21}), the mass loss  behaves the same for these softenings.}
\label{app12}
\end{figure}

\begin{figure}
\centering
\includegraphics[width=0.47\textwidth]{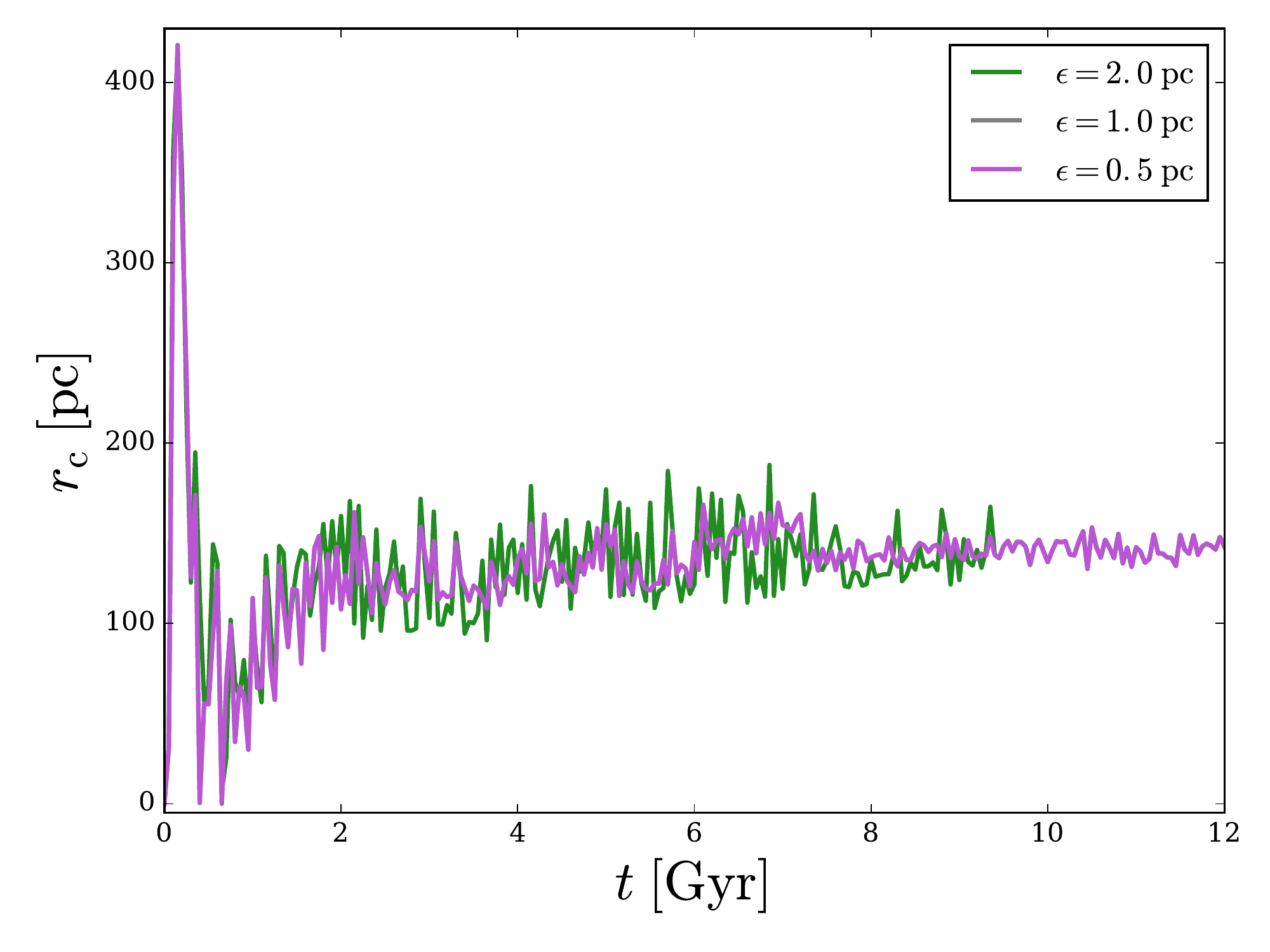}
\caption{{\it Impact of softening on DM halo core size:} Fitted DM core radius as a function of time for three different softening lengths. The core formation is due to crossings of GCs with a DM minihalo in the early accretion scenario. We observe a deviation of the core radius for a softening of 2 pc compared to the other softening lengths. For 1 and 0.5 pc, we noticed that the softening length does not affect the DM core radius. As the stellar density profiles for $\epsilon=0.5$ and 1 pc are nearly identical (see Fig.~\ref{app21}), the DM halo core size behaves the same for these softenings.}
\label{app2}
\end{figure}

In this section, we assess the impact of a numerical parameter that controls the accuracy of our simulations; the softening length $\epsilon$. To test how the softening length impacts on the stellar density profile for isolated GCs with and without a DM minihalo, and the orbital radius and the mass loss of GCs with a DM minihalo and the Fornax DM density profile, we ran simulations with three different softening lengths, $\epsilon$ = 2, 1 and 0.5 pc, in order to ensure that our simulations do not suffer from numerical noise. We apply these tests to the simulation over 12 Gyr in the early accretion scenario for GCs with a DM minihalo. 

As our softening length is similar to the stellar core radius of the GCs, we studied the impact of the softening on the evolution of the stellar density profile over time in Fig.~\ref{app21} for an isolated GC with and without a DM minihalo. Initially, GC stars assume a King profile with a King radius $r_{\rm k} = 1$ pc. The stellar distribution is divided in bins of groups composed of $N_{\rm g}=256$ particles. Fig.~\ref{app21} shows that our mass resolution does not allow us to resolve the 1 pc core radius of GCs. We noticed that the dynamics of GCs is subject to numerical effects for $\epsilon=2$ pc. The convergence of the stellar density profiles confirm that it is sufficient to consider a softening length of 1 pc, which is similar to the King radius for our study.

The evolution of the orbital radius of the five clusters in an accretion process with eccentric orbits (see objects $E_i$ in Table~\ref{tab1}) is shown in Fig.~\ref{app1} for three different softening lengths. It can be seen that the orbital decays of GCs are very similar for all the softening lengths. The evolution of the mass loss for the GC stellar component embedded in a DM minihalo is also providing in Fig.~\ref{app12} for different softening lenghts. Our simulations are well converged for $\epsilon=0.5$ and 1 pc. However, for $\epsilon=2$ pc, numerical noise causes enhanced disruption of the clusters as in Fig.~\ref{app21}. 

In order to determine how the DM density profile depends on the softening length, Fig.~\ref{app2} presents the evolution of fitted DM core radius over time for three different softening lengths. We observe a deviation of the core radius for a softening of 2 pc compared to the other softening lengths. For 1 and 0.5 pc, we deduce that the softening length do not affect the DM core radius. As our simulations with a numerical parameter of 1 and 0.5 pc are well converged over 12 Gyr, we chose $\epsilon$ = 1 pc as the softening length for all our simulations and ensured that our simulations do not suffer from numerical noise.

%%%%%%%%%%%%%%%%%%%%%%%%%%%%%%%%%%%%%%%%%%%%%%%%%%

% Don't change these lines
\bsp	% typesetting comment
\label{lastpage}
\end{document}